\documentclass{aa}
\usepackage{epsfig}
\usepackage{graphicx}
\usepackage[latin1]{inputenc}
\usepackage{amssymb}
\usepackage{amsmath}
\usepackage{relsize}
\usepackage{paralist}
\def\lesssim{\mathrel{\hbox{\rlap{\hbox{\lower4pt\hbox{$\sim$}}}\hbox{$<$}}}}
\def\gtrsim{\mathrel{\hbox{\rlap{\hbox{\lower4pt\hbox{$\sim$}}}\hbox{$>$}}}}
\def\Tex{$T_{\rm ex}$}
\def\mTex{T_{\rm ex}}
\def\abeta{$\bar\beta$}

\def\pbeta{$\beta_\perp$}
\def\ptau{$\tau_\perp$}

\begin{document}
	\title{SiO line emission from C-type shock waves: interstellar jets and outflows}
	\author{A. Gusdorf 
				\inst{1,2} 
				\and S. Cabrit 
				\inst{3} 
				\and D. R. Flower 
				\inst{1} 
				\and G. Pineau des For\^{e}ts 
				\inst{2,3,4}}

\institute{Physics Department, The University, Durham DH1 3LE, UK
\and       Institut d'Astrophysique Spatiale (IAS), B\^{a}timent 121, F-91405 Orsay, France
\and LERMA (UMR 8112 du CNRS), Observatoire de Paris, 61 Avenue de l'Observatoire, F-75014 Paris, France
\and       Universit\'{e} Paris-Sud 11 and CNRS (UMR 8617)}

	\date{received 23 January 2007 / Accepted 01 April 2007}

	\abstract{We study the production of SiO in the gas phase of molecular outflows, through the sputtering of Si--bearing material in refractory grain cores, which are taken to be olivine; we calculate also the rotational line spectrum of the SiO. The sputtering is driven by neutral particle impact on charged grains, in steady--state C-type shock waves, at the speed of ambipolar diffusion. The emission of the SiO molecule is calculated by means of an LVG code. A grid of models, with shock speeds in the range $20 < v_{\rm s} < 50$~km~s$^{-1}$ and preshock gas densities $10^4 < n_{\rm H} < 10^6$~cm$^{-3}$, has been generated. We compare our results with those of an earlier study (Schilke et al. 1997). Improvements in the treatment of the coupling between the charged grains and the neutral fluid lead to narrower shock waves and lower fractions of Si ($\lesssim 10$\%) being released into the gas phase. Erosion of grain cores is significant ($\gtrsim 1$\%) only for C-type shock speeds $v_{\rm s} > 25$~km~s$^{-1}$, given the adopted properties of olivine. More realistic assumptions
concerning the initial fractional abundance of O$_2$ lead to SiO formation being delayed, so that it occurs in the cool, dense postshock flow. Good agreement is obtained with recent observations of SiO line intensities in the L1157 and L1448 molecular outflows. The inferred temperature, opacity, and SiO column
density in the emission region differ significantly from those
estimated by means of LVG `slab' models. The fractional 
abundance of SiO is deduced and found to be in the range $4\times 10^{-8} \lesssim n({\rm SiO})/n_{\rm H} \lesssim 3\times 10^{-7}$. Observed line
profiles are wider than predicted and imply multiple, unresolved
shock regions within the beam.

}

\maketitle

\section{Introduction}
\label{sec:intro}

Unlike CO, which is observed extensively in the interstellar medium of our own and other galaxies, its homologue SiO is observed principally in outflows associated with regions of star formation. This striking difference in behaviour is a consequence of the lower elemental abundance and the more complete depletion of silicon from the gas phase during grain formation. Both carbon and silicon form refractory materials -- graphite and silicates -- of which the cores of interstellar grains are believed to be composed; but the much higher elemental abundance of carbon, $n_{\rm C}/n_{\rm H} = 3.55\times 10^{-4}$, compared with silicon, $n_{\rm Si}/n_{\rm H} = 3.37\times 10^{-5}$ (Anders \& Grevesse 1989), leads to some of the carbon remaining in the gas phase.

The SiO molecule was first detected in the Galactic centre (Sgr B2) by Wilson et al. (1971) and subsequently in Ori A by Dickinson (1972). More recent observations of SiO in jets (see, for example, Bachiller et al. 1991, Mart\'{i}n-Pintado et al. 1992, Codella et al. 1999, Nisini et al. 2007) imply that, in these objects, at least some of the silicon has been restored to the gas phase; this can be achieved through sputtering of the grain material, probably in shock waves, which are features of the outflows. It has been known since the study by \cite{draine} that grain ice--mantles can be eroded in C-type shock waves, owing to impact of neutral particles on charged grains at the ion--neutral drift speed, which is the speed of ambipolar diffusion. More recent work (Flower \& Pineau des For\^{e}ts 1995, Flower et al. 1996, Field et al. 1997, May et al. 2000) has shown that this process might result also in the partial erosion of the refractory grain cores. The simulations undertaken by May et al. of the sputtering of various silicates (forsterite, fayalite and olivine) by neutral atoms showed that C-shock speeds in excess of 30 km s$^{-1}$ are necessary to erode a significant fraction (more than a few per cent) of these materials. On the other hand, it has since been recognized (Le Bourlot et al. 2002, Ciolek et al. 2004) that the speeds at which C-type shock waves can propagate are limited, both by the collisional dissociation of molecular hydrogen, which leads to a thermal runaway and the formation of a J-type shock wave, and by the ion magnetosonic speed, whose value is constrained by the inertia of the charged grains. Consequently, the efficacy of the erosion of silicates in C-type shock waves is restricted by the maximum C-shock speed, which depends on the preshock density and transverse magnetic field strength, and on the fraction of charged grains (Flower \& Pineau des For\^{e}ts 2003).

In a previous study of SiO production in the interstellar medium, \cite{schilke} (henceforth Sch97) considered in detail the erosion of silicon from grain cores and from their mantles by C-type shock waves and the resulting SiO emission spectrum; this study remains the only one of its kind that has been published to date. In the intervening decade, there has been progress in our understanding of the sputtering process (May et al. 2000), the gas-phase chemistry of silicon (Le Picard et al. 2001), and the
grain dynamics (Flower \& Pineau des For\^{e}ts 2003), and so it seems timely to revisit this topic. Recent observations of SiO in jets (Nisini et al. 2007) extend to higher rotational levels than previously and provide a further motivation for such an study. Accordingly, we have reconsidered the chemistry of silicon in steady--state, planar MHD shock waves, with a view to providing a grid of models which may serve in the analysis of current and future observations of rotational transitions of SiO in outflows. These models yield additional results relating to rovibrational transitions of H$_2$ and rotational lines of CO, as well as forbidden lines of atoms and atomic ions, including [Fe II] (cf. Giannini et al. 2006). 

Grain--grain collisions and the sputtering of silicon in oblique C-type shock waves, in which the preshock magnetic field direction is inclined to the plane of the shock front, have been considered by Caselli et al. (1997). However, such simulations have yet to incorporate the gas--phase chemistry and the radiative cooling processes, in order to enable quantitative analyses of the spectral line observations to be made.

\section{The model}
\label{sec:model}

We summarize below those developments, in both the MHD code and the data used and produced by the code, which are relevant to the modelling of the intensities and the profiles of the rotational lines of SiO. We take as our baseline the study by Sch97. The reader is referred to the more recent papers cited below for details of the modifications. 

\subsection{The dynamics of charged grains}

The main revision of the code itself concerns the treatment of the charge and of the dynamical effects of the grains. The charge distribution of both the grains and the ``very small grains'' (VSG), represented by polycyclic aromatic hydrocarbons in our model, are calculated, assuming that collisions with gas--phase particles (electrons, ions and neutrals) determine the charge; see \cite{flower03}. As was mentioned in the Introduction, allowance for the mass density of the (mainly negatively) charged grains can reduce significantly the magnetosonic speed in the charged fluid

\begin{displaymath}
c_{\rm m}^2 = \frac {5 k_{\rm B}(T_+ + T_-)} {3(\mu _+ + \mu _-)} + \frac {B^2}{4\pi (\rho _+ + \rho _-)} 
\end{displaymath}
where $T_{\pm }$ are the temperatures, $\mu _{\pm }$ are the mean masses and $\rho _{\pm }$ are the mass densities of the positively and negatively charged fluids; $B$ is the magnetic field strength in the preshock gas. The magnetosonic speed is the maximum speed at which a C-type shock can propagate in the medium. Furthermore, the momentum transfer between the charged grains and the neutral fluid affects the ion--neutral drift speed and has consequences for the degree of sputtering of the grains within a C-type shock wave.

\subsection{Radiative cooling by H$_2$}

The thermal balance of the medium, particularly the radiative cooling by H$_2$, is treated more exactly in the current model. Rovibrational excitation of H$_2$, principally by H, H$_2$ itself, and He, followed by radiative decay, is the principal mechanism of cooling the shock--heated gas. Following \cite{lebourlot}, the populations of the rovibrational levels of H$_2$ are calculated in parallel with the hydrodynamical and chemical rate equations, yielding the most accurate determination of the rate of cooling by H$_2$ that is achievable within the framework of a time--independent (steady--state) model of the shock structure. The rate coefficients for the collisional excitation by H of rovibrational transitions of H$_2$ are from the recent work of \cite{wrathmall}. Collisional dissociation and ionization of H$_{2}$, as well as ionization of H, are taken into account. Collisional dissociation of H$_2$ is a particularly important process, as the removal of H$_2$ can lead to a thermal runaway. The associated increase in the kinetic temperature, $T_{\rm n}$, of the neutral fluid, and hence of the adiabatic sound speed 

\begin{displaymath}
c_{\rm s}^2 = \frac {5 k_{\rm B}T_{\rm n}}{3\mu _{\rm n}} 
\end{displaymath}
where $\mu _{\rm n}$ is the mean mass of the neutral fluid, can give rise to a sonic point in the flow and hence to a J-type discontinuity. This phenomenon imposes an additional constraint on the maximum speed of a C-type shock wave in a molecular medium.

In the compressed gas of the postshock region where most of SiO forms, cooling by $^{12}$CO, $^{13}$CO, and H$_2$O starts to dominate that by H$_2$. In order to predict accurately the emitted SiO emission spectrum, an LVG treatment of the cooling by these species has been introduced, following the procedures of \cite{neufeld}. A ``thermal gradient'' $c_{\rm s}/z^{\prime }$, where $z^{\prime } = z + 1.0\times 10^{13}$~cm and $z$ is the independent integration variable, is added quadratically to the macroscopic velocity gradient, in order to simulate photon escape through thermal line broadening.

\subsection{The sputtering of grains}

The sputtering probabilities computed by May et al. (2000) for olivine (MgFeSiO$_4$) are used to determine the rate of erosion of Si from (charged) silicate grains by neutral particles. We include also the sputtering of carbonaceous (amorphous carbon) grains, using the sputtering yields of \cite{field}. Impacts (at the ion--neutral drift speed) of abundant heavy neutral species are the most effective in eroding the grain cores. For collisions with CO, for example, we adopted the sputtering probabilities calculated for impacts of Si atoms, which have the same mass as CO and hence the same impact energy. As \cite{may} have shown, similar yields of Si are obtained from the three types of silicate: fayalite, Fe$_2$SiO$_4$; forsterite, Mg$_2$SiO$_4$; and olivine, MgFeSiO$_4$. 

The grain mantles are eroded first, through impacts with the most abundant neutral species, H, H$_2$ and He, at ion--neutral drift speeds which are below the threshold for erosion of the cores (Draine et al. 1983, Flower \& Pineau des For\^{e}ts 1994). The initial composition of the gas is calculated in chemical equilibrium, whilst that of the grain mantles, and the elemental depletion into the grain cores, is based on observations (cf.~Flower \& Pineau des For\^{e}ts 2003, table~1). We incorporated a representative polycyclic aromatic hydrocarbon (PAH), with a fractional abundance $n_{\rm PAH}/n_{\rm H} = 10^{-6}$, an upper limit which is consistent with estimates of the fraction of elemental carbon likely to be present in the form of very small grains (Li \& Draine 2001; Weingartner \& Draine 2001). The state of charge of this species was computed following \cite {flower03}, who showed that increasing the fractional abundance of the PAH results in higher values of the magnetosonic speed in the preshock gas, thereby enabling C-type shock waves to propagate at higher speeds. In their turn, the higher speed shocks erode the silicate grains more effectively. However, the adopted
PAH abundance has little effect on the internal structure of the shock wave, as the grains become negatively charged in the shock wave and dominate
grain--neutral momentum transfer.

The total number density of the grains was computed 

\begin{itemize}

\item from the total mass of the refractory cores, which is $0.78\times 10^{-2}$ times the mass of the gas for the adopted composition of the cores; 

\item assuming a bulk mass density of 3~g~cm$^{-3}$; and 

\item taking a size distribution of the grain cores $${\rm d}n_{\rm g}(a_{\rm g})/{\rm d}a_{\rm g} \propto a_{\rm g}^{-3.5},$$ following \cite{mathis77}, where $n_{\rm g}(a_{\rm g})$ is the number density of grains with radius $a_{\rm g}$, assumed to have upper and lower limits of 0.3~$\mu $m and 0.01~$\mu $m, respectively. 

\end{itemize}

The thickness of the grain mantles was determined from their molecular composition (Flower \& Pineau des For\^{e}ts 2003, table~2), assuming a mean number of $5\times 10^4$ molecular binding sites per layer of the mantle and a thickness of $2\times 10^{-4}$~$\mu $m for each layer, independent of the size of the grain core; there are no Si--containing species in the mantles. This procedure yields an initial mantle thickness of $0.015$~$\mu $m on the grain cores, whose mean radius is $a_{\rm g} = 0.02$~$\mu $m. However, the erosion of the grain mantles occurs sufficiently rapidly, as the ion and neutral flow velocities begin to decouple in the shock wave, that the existence of thick ice--mantles on the grains in the preshock gas has little effect on the shock dynamics (see fig.~6 of Guillet et al. 2007). 

\subsection{Chemistry}
\label{sec:chemistry}

The chemical reaction network comprises over 900 reactions connecting the abundances of more than 100 species, in both the gas and the solid phases. The complete list of species and reactions is available from http://massey.dur.ac.uk/drf/outflows\_test/species\_chemistry\_shock/. In the context of the present study, we emphasize the gas--phase chemistry of silicon and the formation of SiO, in particular. The total rate of cosmic ray ionization of hydrogen, $\zeta $, was taken to be $5\times 10^{-17}$~s$^{-1}$.

Two reactions are important for the formation of SiO from Si, which is released into the gas phase by the sputtering of the grains, namely \\
\begin{equation}
\begin{array}{ccc}
\label{eqs1}
\rm {Si} + \rm {O}_{2} & \longrightarrow  & \rm {SiO} + \rm {O}\\
\end{array}
\end{equation}
\begin{equation}
\begin{array}{ccc}
\label{eqs2}
\rm {Si} + \rm{OH}  & \longrightarrow  & \rm {SiO} + \rm {H} \\
\end{array}
\end{equation}
for which the following rate coefficients (cm$^3$~s$^{-1}$)

\begin{equation}
\begin{array}{ccc}
\label{eqs3}
k_1 & = & 2.7\times 10^{-10} \exp({-111/T}) \\
\end{array}
\end{equation}
\begin{equation}
\begin{array}{ccc}
\label{eqs4}
k_2 & = & 1.0\times 10^{-10} \exp({-111/T}) \\
\end{array}
\end{equation}
were adopted by Sch97 from the compilation of \cite{langer}. The exponential factor in equation~(\ref{eqs3}) derives from the argument (Graff 1989) that the reactions proceed only with the fine--structure states of Si~(3p$^2$ $^3$P$_J$) with $J > 0$, of which $J = 1$, which lies 111~K above the $J = 0$ ground state, is the more significantly populated at low temperatures. 
More recently, the rate coefficient for reaction~(\ref{eqs1}) has been measured at low temperatures ($15 \le T \le 300$~K) by \cite{lepicard} and found to be given by 

\begin{equation}
\label{eqs5}
k_1 = 1.72\times 10^{-10} (T/300)^{-0.53} \exp(-17/T).
\end{equation}
We have adopted~(\ref{eqs5}) and the same expression for $k_2$. Evidently, the differences between the present and previous values of these rate coefficients are most significant for temperatures $T \lesssim 100$~K, i.e. in the cooling flow of the shocked gas.

The abundance of SiO is limited by its conversion to SiO$_2$ in the reaction with OH

\begin{equation}
\begin{array}{ccc}
\label{eqs6}
\rm {SiO} + \rm{OH}  & \longrightarrow  & \rm {SiO_2} + \rm {H} \\
\end{array}
\end{equation}
whose rate coefficient remains subject to considerable uncertainty. We adopt the same expression as Sch97, viz.

\begin{equation}
\label{eqs7}
k_6 = 1.0\times 10^{-11} (T/300)^{-0.7} 
\end{equation}
in units of cm$^3$~s$^{-1}$. However, we note that \cite{zach} calculated a barrier of 433~K to reaction~(\ref{eqs6}), and a rate coefficient $$k_6 = 2.5\times 10^{-12} (T/300)^{0.78}\exp(-613/T);$$
see the discussion of \cite{lepicard}. At $T = 300$~K, the latter rate coefficient is 30 times smaller than the former. In the ambient (preshock) and the postshock gas, where $T \approx 10$~K, the existence of an activation energy of several hundred kelvin would prevent the oxidation of SiO in reaction~(\ref{eqs6}) from occurring. The rate coefficient for reaction~(\ref{eqs6}) in the UMIST data base (Le Teuff et al. 2000) is $$k_6 = 2.0\times 10^{-12}$$ in cm$^3$~s$^{-1}$, which is 50 times smaller than (\ref{eqs7}) at $T = 10$~K. Fortunately, the conversion of SiO into SiO$_2$ occurs in a region which is too cold and optically thick to contribute to the SiO line intensities, and so the uncertainty in the rate coefficient for reaction~\ref{eqs6} is not significant in the present context.

Re-adsorption of molecules on to grains (``freeze--out'') in the postshock gas is included, as in Sch97. The effects of freeze--out on the abundance of SiO and its spectrum are considered below.

\subsection{Line transfer}
\label{sec:transfer}

The physical and chemical profiles which derive from the shock model summarized above are used in a large velocity gradient (LVG) calculation of the intensities of the emission lines of SiO and of CO. Our implementation of this technique is described in Appendix~\ref{appendix}, where we note that an expression for the escape probability which differs from that of Sch97 has been adopted.

\section{Results}
\label{sec:results}

\subsection{Comparison with the calculations of \cite{schilke}}
\label{sub:comparison}

\begin{figure}
%\resizebox{\hsize}{!}{\includegraphics[angle=-90]{graphes/schilke_T_t.ps}}
%\resizebox{\hsize}{!}{\includegraphics[angle=-90]{graphes/schilke_v_t.ps}}
\includegraphics[height=20cm]{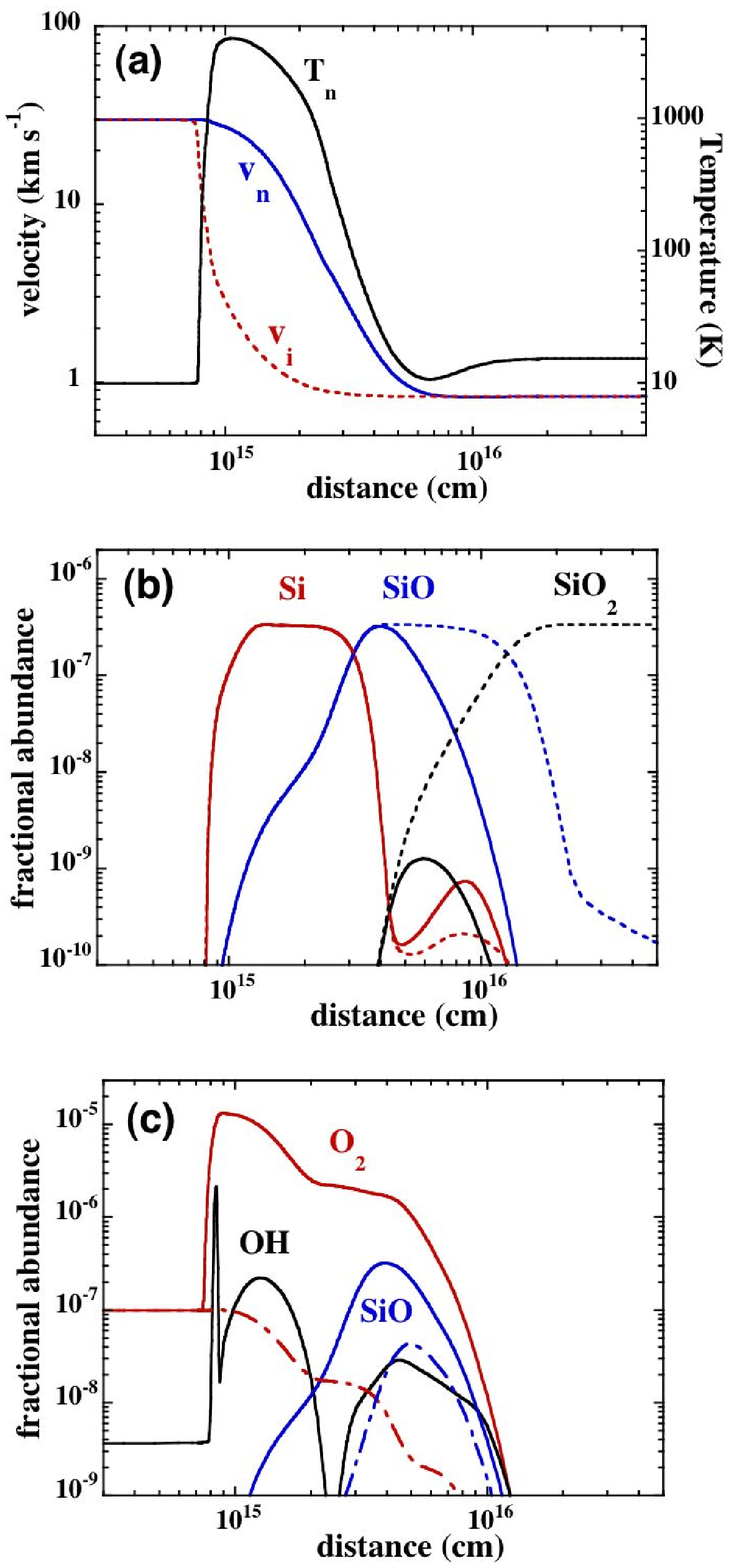}
\caption{(a) Temperature of the neutral fluid and velocity profiles of the neutral and charged fluids, predicted by the present C-type shock model. The shock parameters are  $n_{\rm H} = 10^{5}$~cm$^{-3}$ and $B = 200$~$\mu$G in the preshock gas, and $v_{\rm s} = 30$~km~s$^{-1}$, $\zeta = 5\times 10^{-17}$~s$^{-1}$. The fractional gas--phase abundances of selected Si--and O--bearing species are plotted in panels (b) and (c); cf. Sch97, fig.~2. The broken curves in panel (b) are the corresponding results obtained when re-adsorption on to grains is neglected. The discontinuous curves in panel (c) show the effects of assuming the initial abundance of O$_2$ ice to be negligible, i.e. the second of the two scenarios described in Section~\ref{sub:comparison}.}
\label{schilke_1}
\end{figure}

First, we compare the computed shock structure with that of Sch97, for a reference C-type shock model, in which the preshock density $n_{\rm H} = 10^{5}$~cm$^{-3}$ and the magnetic field strength $B = 200$~$\mu$G, and the shock velocity $v_{\rm s} = 30$~km~s$^{-1}$. The most striking difference between the current and the previous model is that the width of the shock wave decreases by a factor of approximately 4, to $5\times 10^{15}$~cm, from $2\times 10^{16}$~cm in the study of Sch97; see Fig.~\ref{schilke_1}a. This difference is attributable to the more accurate treatment of the coupling between the neutral fluid and the charged grains in the current model and is an indication of the significance of the inertia of the (negatively) charged grains in dark clouds, in which the degree of ionization is low. With the narrower shock wave is associated a higher maximum temperature of the neutral fluid, as there is less time for the initial energy flux, $\rho_{\rm n}v_{\rm s}^3/2$, associated with the bulk flow, to be converted into internal energy of the H$_2$ molecules or to be radiated away. Thus, $T_{\rm n} \approx 4000$~K here, compared with $T_{\rm n} \approx 2000$~K in the model of Sch97. 

There are chemical differences between the models also, which are related in part to the changes in the shock structure; these differences may be summarized as follows.

\begin{compactenum}[(i)]

\item The fraction of the Si in the grain cores which is released into the gas phase by sputtering is approximately ten times smaller in the current model than in the model of Sch97. This reduction is attributable partly to the sputtering yields, which have higher thresholds and are smaller for olivine (MgFeSiO$_4$) than for the amorphous silica (SiO$_2$) considered by Sch97; but the main reason for the decrease is the enhanced coupling between the neutral fluid and the charged grains, which reduces the shock width and hence the time available to erode the grains. On the other hand, the magnitude of the ion--neutral drift speed is similar in both calculations. As a consequence of the reduction in the shock width, the integrated SiO line intensities predicted by the current model are smaller, in general, than calculated by Sch97; see Section~\ref{sec:sio}.

\item The displacement of the maximum fractional abundance of SiO, which forms in the gas--phase reactions~(\ref{eqs1}) and (\ref{eqs2}), from that of Si, which is eroded from the grains, is a more significant fraction of the shock width in the current model; cf. Fig.~\ref{schilke_1}b. The initial fractional abundance of O$_2$ in the preshock medium is lower here, by a factor of approximately 10, than in the model of Sch97, delaying the initial formation of SiO. The O$_2$ is assumed to be predominantly in the form of ice, which is sputtered rapidly from the grains in the early stages of development of the shock wave, as may be seen from the two orders of magnitude increase in the fractional abundance of gas--phase O$_2$, apparent in Fig.~\ref{schilke_1}c. The fractional abundances of O$_{2}$ and OH decrease subsequently, at high kinetic temperatures, owing to their dissociation by H in the chemical reactions O$_2$(H, O)OH and OH(H, O)H$_2$. The former reaction, which is endothermic by over 8000~K, proves to be less effective in destroying O$_2$ over the smaller width of the current shock model (see Fig.~\ref{schilke_1}c) than was the case in the calculations of Sch97. On the other hand, the lower energy threshold of 17~K in reaction~(\ref{eqs5}) allows oxidation reactions to proceed further into the postshock region, compared with Sch97, whose adopted threshold was 111~K. As a consequence, conversion of Si into SiO is slower initially but more complete eventually than predicted by Sch97.

\item SiO$_{2}$ is removed more rapidly from the gas phase in the current model. The compression is more rapid than in the model of Sch97, and so the rate of adsorption of molecules to grains (``freeze--out'') is higher. If the oxidation of SiO in the reaction SiO(OH, H)SiO$_2$ has an activation energy of several hundred kelvin (\cite{zach}; see Section~\ref{sec:chemistry}), the maximum fractional abundance of SiO$_2$ would be reduced still further.

\end{compactenum}

Also shown in Fig.~\ref{schilke_1}b are the fractional abundances which are obtained neglecting re-adsorption on to the grains. The timescale for freeze--out is sufficiently large that the chemical profiles are modified only in the cold postshock gas, where the flow speed is practically constant and the optical depths in the SiO lines are large. Consequently, whilst the effects re-adsorption on the composition of the postshock gas are dramatic, the freeze--out of SiO has no effect on the predicted line intensities.

In chemical equilibrium, the initial fractional abundance of O$_2$ is approximately $10^{-5}$, whereas observations with the Odin satellite (Pagani et al. 2003, Larsson et al. 2007) have placed upper limits of $n({\rm O}_2)/n({\rm H}_2) \lesssim 10^{-7}$. In view of these measurements, we have considered two scenarios, both with an initial gas--phase fractional abundance $n({\rm O}_2)/n_{\rm H} = 1.0\times 10^{-7}$, as a consequence of the freeze--out of oxygen on to grains, but with differing assumptions regarding its chemical form in the grain mantles.

\begin{itemize}

\item The molecular oxygen which formed in the gas phase was adsorbed on to the grains, where it remained as O$_2$ ice in the preshock medium, with a fractional abundance of $1.3\times 10^{-5}$, relative to $n_{\rm H}$. The initial fractional abundance of H$_2$O ice is an order of magnitude larger than that the fractional abundance of O$_2$ ice.

\item Atomic oxygen was adsorbed on to the grains before O$_2$ was synthesized and subsequently hydrogenated to H$_2$O ice in the grain mantles of the cold preshock medium. The fractional abundance of H$_2$O ice increases by only 25\% as a consequence, to $1.3\times 10^{-4}$.

\end{itemize}

It turns out that the first scenario is practically equivalent to assuming that the O$_2$ is initially in the gas--phase (see Fig.~\ref{schilke_1}c), as its release from the grain mantles occurs early and rapidly (on a timescale of a few years for the model in Fig.~\ref{schilke_1}) in the shock wave. On the other hand, the second scenario can result in reduced levels of oxidation of Si to SiO in the gas--phase (cf.  Fig.~\ref{schilke_1}), depending on the relative importance of reactions~\ref{eqs1} and \ref{eqs2} in the oxidation process. In what follows, we present results corresponding principally to the first scenario, with the second being considered mainly in Appendix~\ref{O_2}. 

\subsection{A grid of models}
\label{sub:grid}

We have computed a grid of shock models with the following parameters:

\begin{itemize}

\item $n_{\rm H} = 10^{4}$~cm$^{-3}$, $v_{\rm s} = 20, 25, 30, 35, 40, 45, 50$~km~s$^{-1}$;

\item $n_{\rm H} = 10^{5}$~cm$^{-3}$, $v_{\rm s} = 20, 25, 30, 35, 40, 45$~km~s$^{-1}$;

\item $n_{\rm H} = 10^{6}$~cm$^{-3}$, $v_{\rm s} = 20, 25, 27, 30, 32, 34$~km~s$^{-1}$.

\end{itemize}
In all of these models, we characterized the preshock magnetic field strength by 
$$B = bn_{\rm H}^{0.5},$$ where $n_{\rm H}$ is in cm$^{-3}$ and $B$ is in $\mu $G; the scaling parameter $b$ was taken equal to 1. (The effect on Si sputtering of varying $b$ is discussed in Section~\ref{sub:magnetic}.) The maximum shock speed for $n_{\rm H} \gtrsim 10^4$~cm$^{-3}$ is determined by the collisional dissociation of H$_2$, the main coolant, which leads to a thermal runaway and a J-discontinuity (Le Bourlot et al. 2002, Flower \& Pineau des For\^{e}ts 2003). 

In fact, we computed two grids, one for each of the scenarios concerning the initial distribution of oxygen between O$_2$ and H$_2$O ices, as specified towards the end of the previous Section~\ref{sub:comparison}. We concentrate on the first of these two scenarios, but some additional Figures for the
second scenario are given in Appendix~\ref{O_2} as online material. (Our results are available in digital tabular format on request to the authors.)

\begin{figure}
\includegraphics[height=20cm]{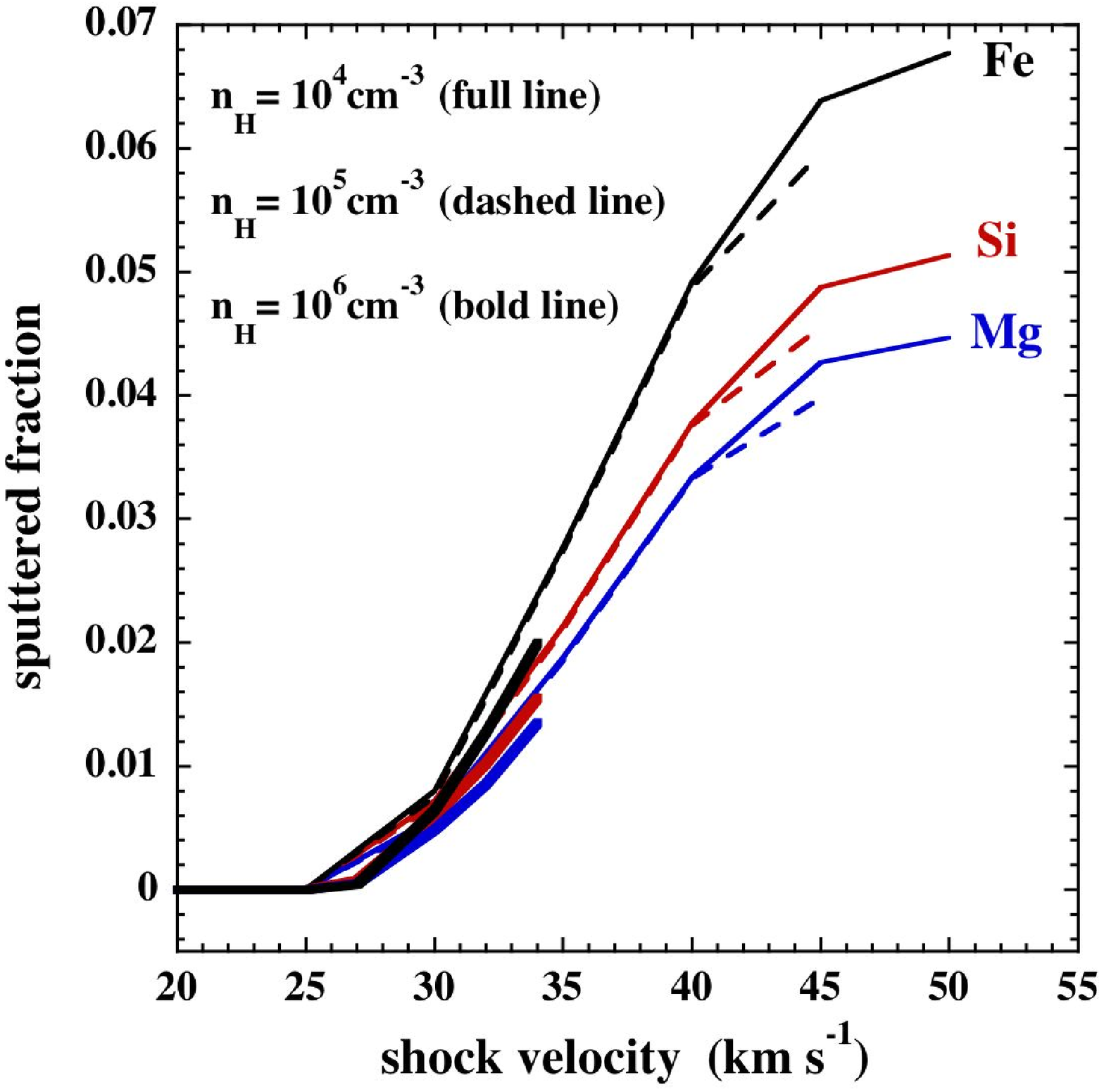}
\caption{The fractions of Mg, Si and Fe, initially in the form of olivine (MgFeSiO$_4$), which are released into the gas phase by sputtering within a steady--state C-type shock wave.}
\label{si_1}
\end{figure}

Because of the sharply defined sputtering threshold energy of approximately 50~eV, there is negligible sputtering of Si from the olivine (MgFeSiO$_4$) for shock speeds of 20~km~s$^{-1}$ or less. The fractions of the Mg, Si and Fe which are released from the olivine into the gas phase are shown in Fig.~\ref{si_1}. Comparing Fig.~\ref{si_1} with the corresponding figure~4 of \cite{may}, whose sputtering yields are used in the present calculations, shows that the fractions of Mg, Si and Fe which are sputtered from olivine have decreased by an order of magnitude. As the same sputtering yields have been used in both studies, this change is attributable to the reduction in the shock width, resulting from the improved treatment of grain--neutral coupling. We note that CO is the principal eroding partner (cf. May et al. 2000). 

Fig.~\ref{si_1} shows that the degree of sputtering is, in fact, insensitive to the preshock gas density (cf. Caselli et al. 1997); it depends essentially on the shock speed. Polynomial fits of the sputtered fractions of Fe, Si and Mg, as functions of the shock speed, are given in Appendix~\ref{appendix_bis}.

In Fig.~\ref{si_sio}, we display the fractional gas--phase abundances of Si and SiO, as functions of the relevant flow time. Silicon is produced by erosion of the charged grains by collisions, principally with molecules, at the ion--neutral drift speed. Once the drift speed exceeds the sputtering threshold velocity, the erosion of Si occurs rapidly, as Fig.~\ref{si_sio} shows. Thus, the flow time which is directly relevant to the release of Si into the gas phase is that of the {\it charged} fluid, rather than that of the neutrals, which is the appropriate measure of the total time for formation of SiO. As noted in item~(ii) of Section~\ref{sub:comparison}, there is an additional, chemical delay to the conversion of Si into SiO, in reactions~(\ref{eqs1}) and (\ref{eqs2}), which is apparent in our Fig.~\ref{si_sio}, owing to the low abundance and partial destruction of O$_2$. The magnitude of this delay depends on the parameters of the model, notably the shock speed, $v_{\rm s}$, and the preshock gas density, $n_{\rm H}$. Conversion is almost instantaneous for $v_{\rm s} \ge 30$~km~s$^{-1}$, $n_{\rm H} = 10^6$~cm$^{-3}$, when OH is formed abundantly at the start of the shock and reaction~(\ref{eqs2}) dominates the oxidation process.

Fig.~\ref{sio_1} shows the variation with shock speed and preshock gas density of the fractional abundance of SiO, computed through the entire shock wave. It is evident from Fig.~\ref{sio_1} that the duration of the C-type shock wave, as measured by the temperature profile, is of the order of $10^4$, $10^3$ and $10^2$~yr for preshock gas densities $n_{\rm H} = 10^4$, $10^5$ and $10^6$~cm$^{-3}$, respectively. The peak SiO abundance is reached over similar timescales. It may be seen that the highest fractional abundances of SiO are attained for the lowest preshock density, $n_{\rm H} = 10^4$~cm$^{-3}$. At higher densities, both O$_2$ and OH, which are reactants in (\ref{eqs1}) and (\ref{eqs2}), are destroyed by the atomic hydrogen which is produced in the shock wave. Thus, the conversion of Si into SiO
becomes incomplete at high density, and the gas--phase SiO abundance depends on $n_{\rm H}$ even though the Si sputtered fraction does not (cf. Fig.~\ref{si_1}).

\begin{figure}
\includegraphics[height=20cm]{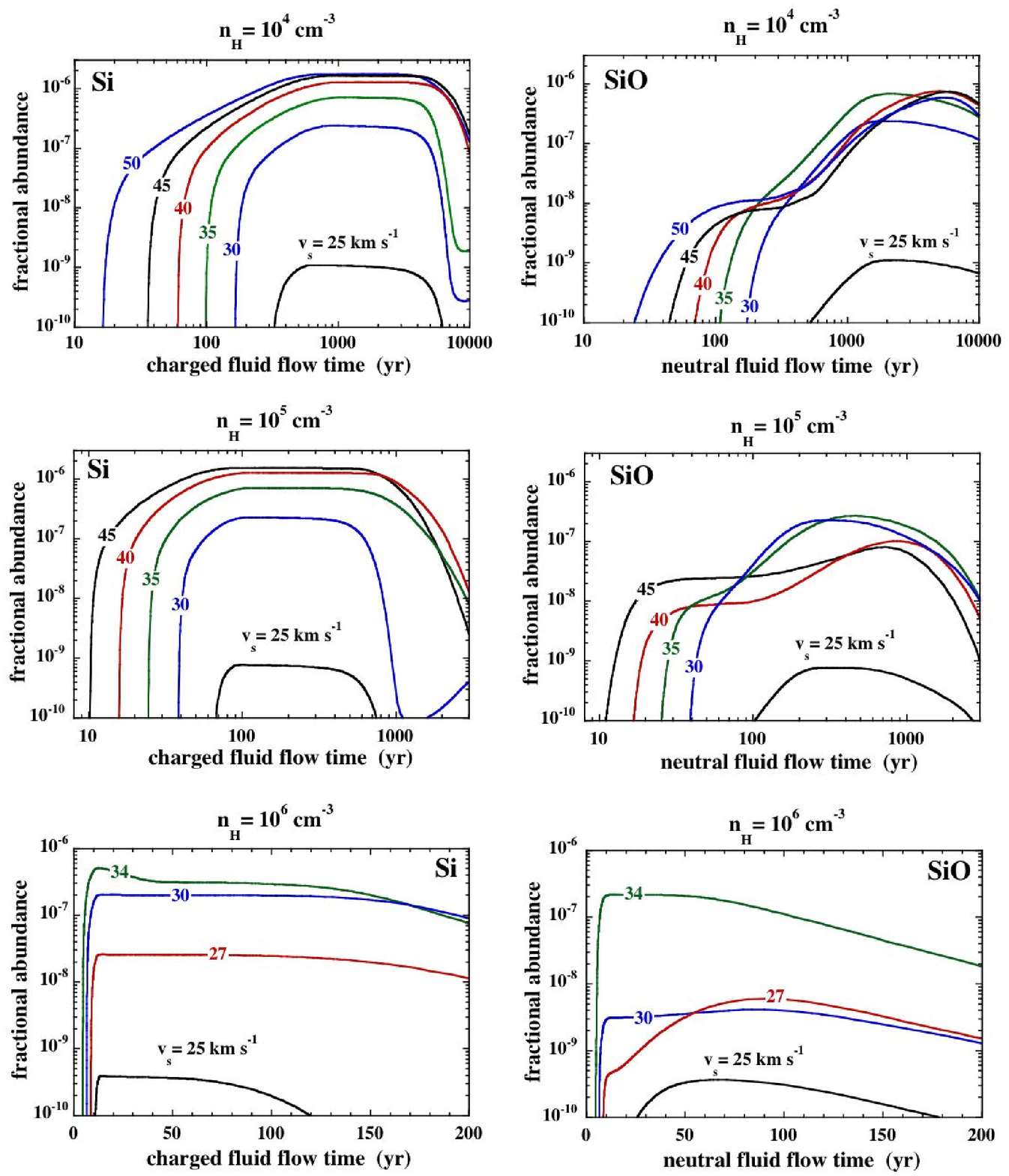}
\caption{The fractional abundances of Si, released into the gas phase by the sputtering of olivine (MgFeSiO$_4$), and of SiO, which subsequently forms in reactions~(\ref{eqs1}) and (\ref{eqs2}). In the left--hand panels, the independent variable (abscissa) is the flow time of the charged fluid: see text, Section~\ref{sub:grid}.}
\label{si_sio}
\end{figure}

\begin{figure}
\includegraphics[height=20cm]{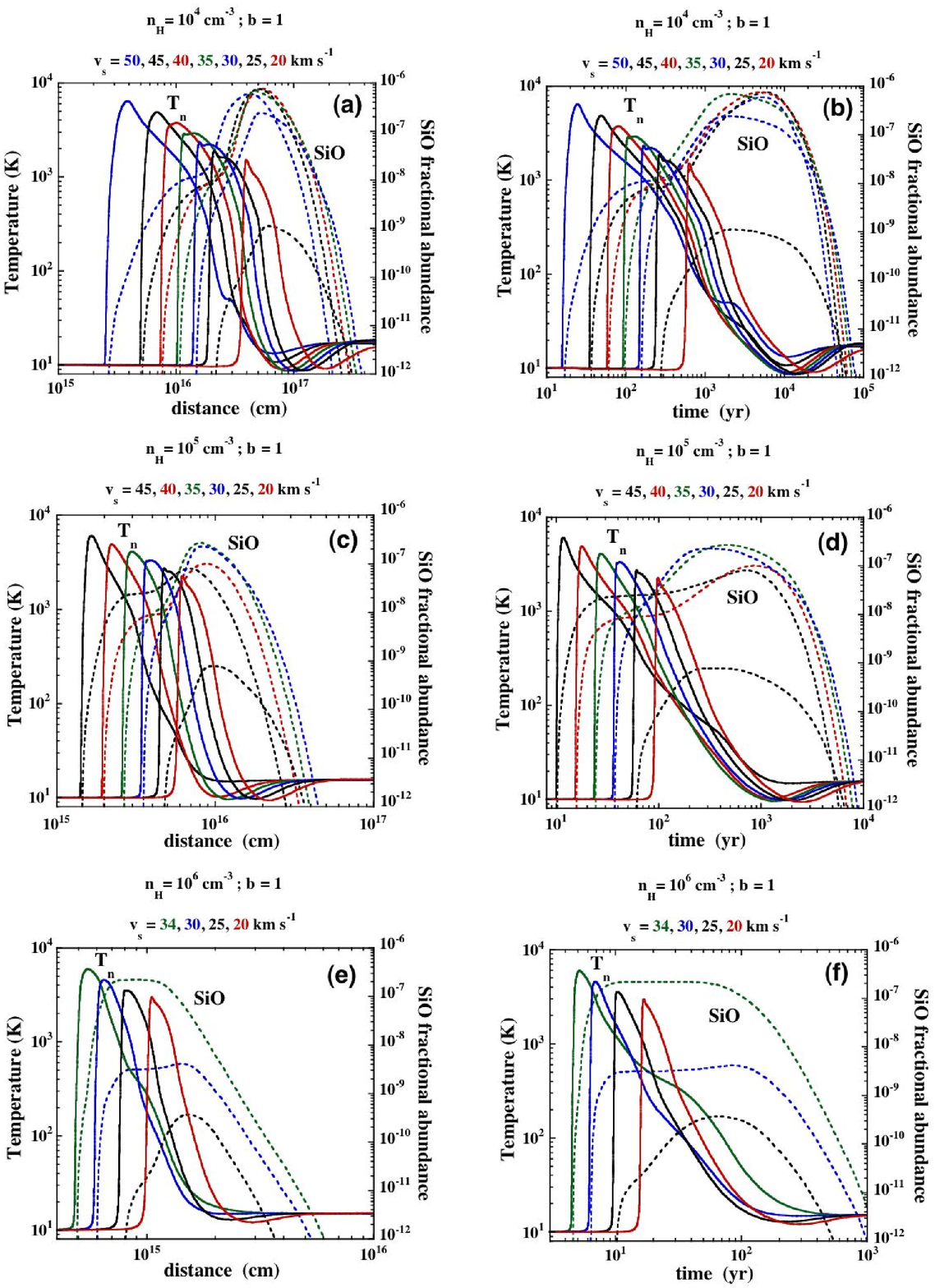}
\caption{The fractional abundance of SiO, $n$(SiO)/$n_{\rm H}$, computed for the grid of shock models and plotted as a function of distance (left) and neutral
flow time (right); $n({\rm SiO})/n_{\rm H}$ is negligible for $v_{\rm s} \lesssim 20$~km~s$^{-1}$. In addition, the temperature of the neutral fluid is plotted. (See also Fig.~\ref{O2_4} of Appendix~\ref{O_2}.)}
\label{sio_1}
\end{figure}

\subsection{Dependence on the transverse magnetic field strength}
\label{sub:magnetic}

The existence of a magnetic field transverse to the direction of propagation is a necessary condition for a C-type shock wave to form, and it is instructive to consider the variation of the structure of the shock wave with the strength of the magnetic field. Energy equipartition arguments, applied to the magnetic and thermal energy densities in the preshock molecular gas, of particle density $n({\rm H}_2) + n({\rm He}) = 0.6n_{\rm H}$ and kinetic temperature $T$, suggest that $B^2/(8\pi ) \approx n_{\rm H}k_{\rm B}T$ and hence that $B = bn_{\rm H}^{0.5}$, where $b$ is a scaling parameter (cf. Section~\ref{sub:grid}) such that $B$ is in $\mu $G when $n_{\rm H}$ is in cm$^{-3}$; this is the proportionality adopted in the grid of models presented in Section~\ref{sub:grid}. In gas of $T = 10$~K, equipartition with the thermal energy implies $b = 0.18$. However, we note that such a low value of $b$ is inconsistent with the existence of a steady--state C-type shock wave when $n_{\rm H} = 10^{5}$~cm$^{-3}$ and $v_{\rm s} \ge 10$~km~s$^{-1}$, as the corresponding ion magnetosonic speed in the preshock gas (9.7~km~s$^{-1}$) is lower than the shock speed.

\begin{figure}
\includegraphics[height=20cm]{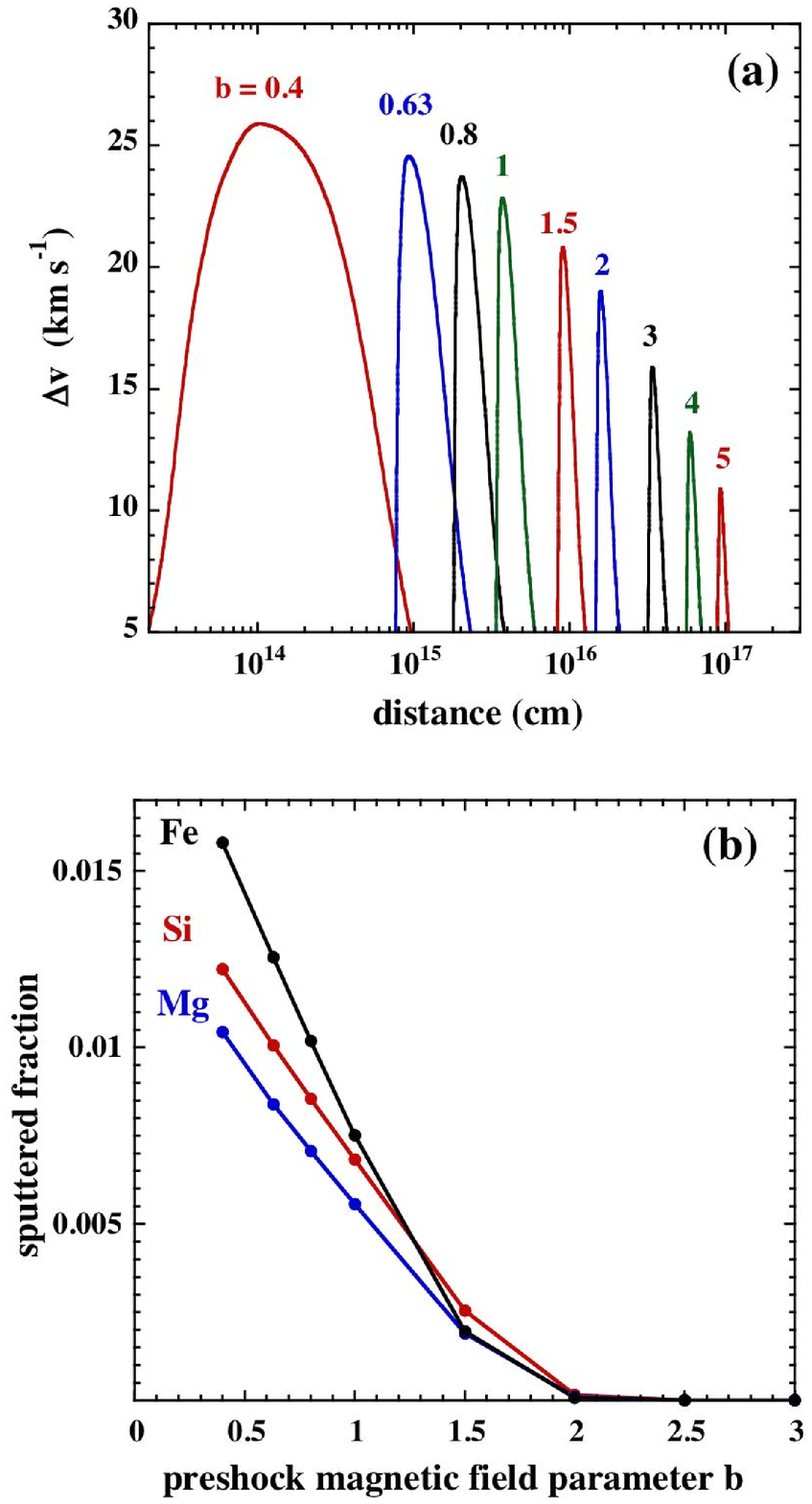}
\caption{(a) The ion--neutral velocity difference, $\Delta v = |v_{\rm i} - v_{\rm n}|$, and (b) the fractions of Mg, Si and Fe eroded from olivine (MgFeSiO$_4$) grains, computed as functions of the transverse magnetic field strength, $B$, in the preshock gas; $n_{\rm H} = 10^{5}$~cm$^{-3}$ and $v_{\rm s} = 30$~km~s$^{-1}$. Note that $B = bn_{\rm H}^{0.5}$, where $b$ is a scaling parameter (cf. Section~\ref{sub:grid}) such that $B$ is in $\mu $G when $n_{\rm H}$ is in cm$^{-3}$.}
\label{Bfield}
\end{figure}

In Fig.~\ref{Bfield}, we present results as a function of the transverse magnetic field strength, for the parameters of the model of Sch97: $n_{\rm H} = 10^{5}$~cm$^{-3}$, $v_{\rm s} = 30$~km~s$^{-1}$, and a magnetic field scaling parameter $b$ in the range $0.5 \le b \le 5$. We recall that Sch97 adopted $B = 200$~$\mu$G, corresponding to $b = 0.63$. This value of $b$ is consistent with the analysis of Zeeman measurements by \cite{crutcher}, who concluded that there was an approximate equipartition of the magnetic and kinetic energy densities in the molecular clouds that he had observed. It may be seen from Fig.~\ref{Bfield} that increasing the magnetic field inhibits the release of Si from refractory grain cores in the shock wave, owing to the reduction in the maximum ion--neutral velocity difference, $\Delta v$. In Section~\ref{sub:sio_5}, we consider how the magnetic field strength affects the relative intensities of the rotational transitions of SiO.

\section{SiO rotational emission lines}
\label{sec:sio}

The observable quantities are the intensities of the rotational transitions of SiO and the velocity--profiles of these emission lines. Having computed the shock structure, we evaluate the line intensities and profiles as described in Appendix~\ref{appendix}, assuming that the shock is viewed face--on.

\subsection{Physical conditions in the SiO emission region}
\label{sub:sio_1}

\begin{figure}
\includegraphics[height=20cm]{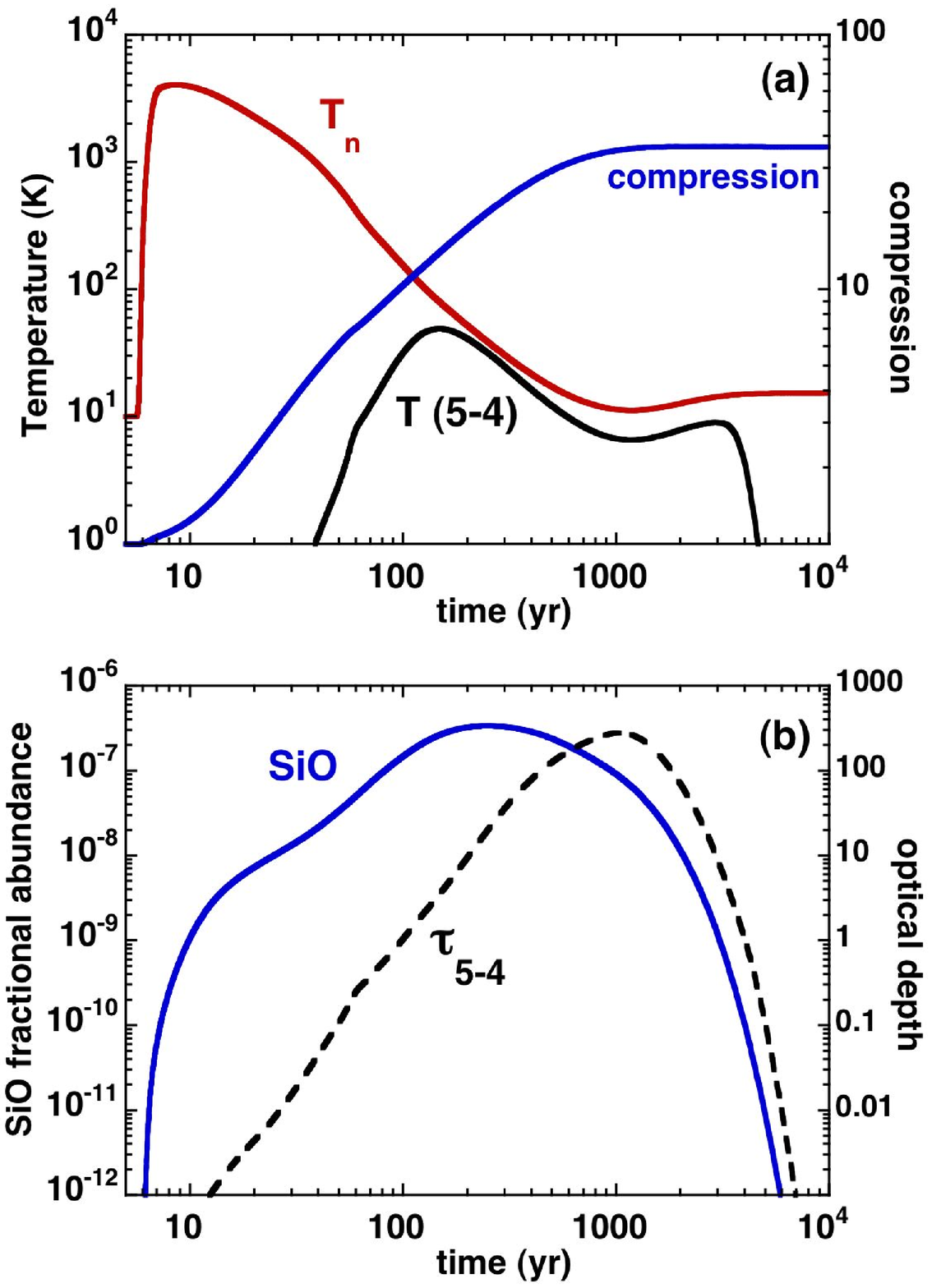}
\caption{(a) The temperature of the neutral fluid, $T_{\rm n}$, the brightness temperature, $T(5-4)$, in the $j =$~5--4 line, and the compression factor, $n_{\rm H}/n_{\rm H}({\rm initial})$; (b) the optical depth, $\tau _{5-4}$ in the 5--4 transition and the fractional abundance of SiO, $n({\rm SiO})/n_{\rm H}$, as functions of the flow time of the neutral fluid, $t_{\rm n}$. The model parameters are $n_{\rm H} = 10^{5}$~cm$^{-3}$, $v_{\rm s} = 30$~km~s$^{-1}$, and $b = 0.63$.}
\label{phys_cond_1}
\end{figure}

Fig.~\ref{phys_cond_1} illustrates the variation of physical conditions throughout
the formation region of the SiO 5--4 rotational line for our reference model
with $n_{\rm H} = 10^{5}$~cm$^{-3}$, $v_{\rm s} = 30$~km~s$^{-1}$, and $b = 0.63$. It may be seen that the line is optically thin through most of the hot precursor (where the flow speeds of the charged and neutral fluids differ), due to both the low
SiO abundance and the large velocity gradient there. Therefore the
line intensity is low despite a high kinetic temperature.
At the rear of the shock wave, approaching maximum compression, the synthesis of SiO and the steady decrease in velocity gradient eventually raise the optical depth in the line, and the 5--4 intensity peaks, with the line temperature attaining values close to the local kinetic temperature of the neutral fluid, $T_{\rm n}$; that is, the line approaches LTE. The intensity then declines rapidly as the gas cools; the
decline occurs in 500~yr for the model shown here. This behaviour is insensitive to
the rate of re-adsorption of SiO on to the grains, which occurs over much longer timescales.

\begin{figure}
\includegraphics[height=20cm]{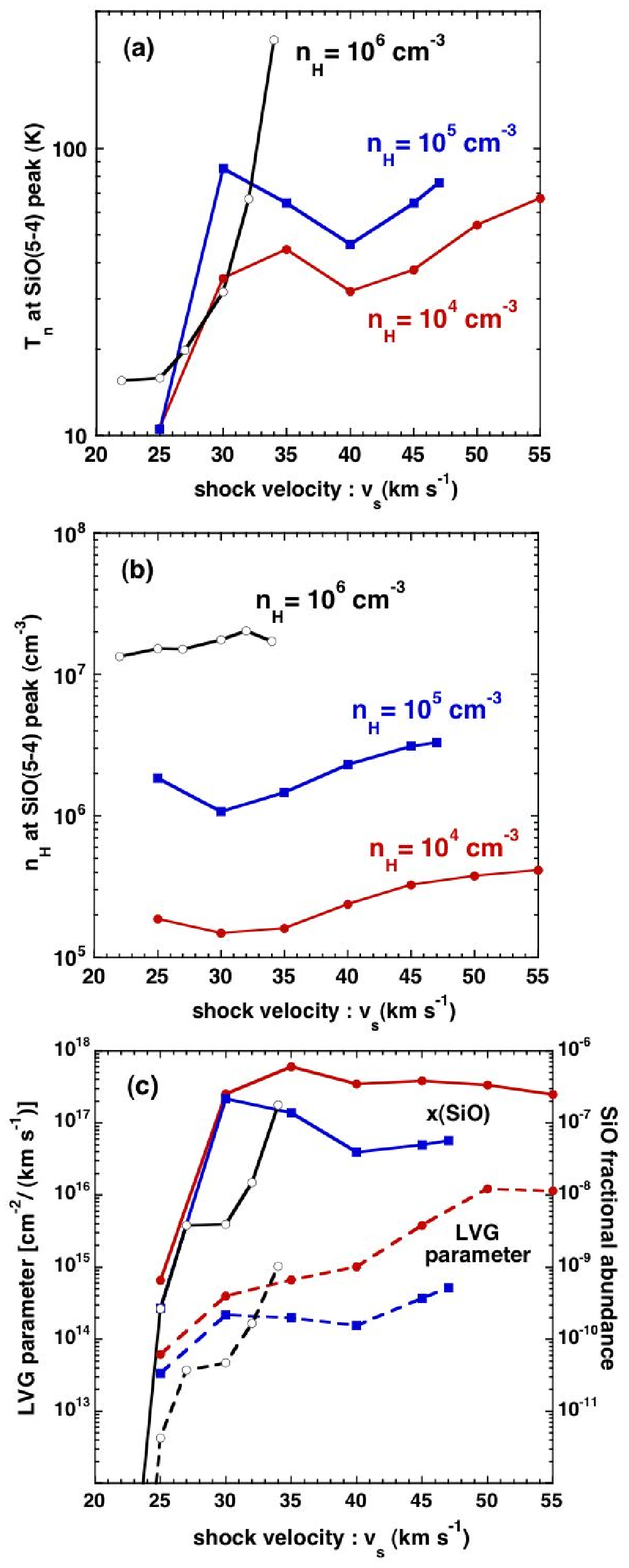}
\caption{Physical conditions at the position of the peak in the SiO 5--4 line intensity [$T_{\rm peak}(5-4)$] as functions of the shock speed, $v_{\rm s}$, 
for all models of the grid: (a) neutral temperature, $T_{\rm n}$; (b) total density, $n_{\rm H}$; (c) the LVG parameter, $n({\rm SiO})/({\rm d}v_z/{\rm d}z)$, and the fractional abundance of SiO, $x({\rm SiO}) \equiv n({\rm SiO})/n_{\rm H}$. (See also Fig.~\ref{O2_7} of Appendix~\ref{O_2}.)}
\label{phys_cond_2}
\end{figure}

In order to illustrate the dependence of conditions in the SiO emission region on the shock parameters, we present, in Fig.~\ref{phys_cond_2}, the important physical
quantities, evaluated {\it at the peak of the SiO 5--4 line}, for all
models in our grid. We have verified that this is equivalent to
computing intensity--weighted geometric means of the same quantities
over the region where the 5--4 line intensity is more than 50\% of its
maximum value.\footnote{Values
evaluated at the peak also differ by less than a factor 2 from the
same parameters evaluated at the ``median'' point where the integrated line intensity, $T{\rm d}V(5-4)$, reaches half of its total.} Hence, it provides a good indication of the mean characteristics
of the region producing the peak of the emission.

Fig.~\ref{phys_cond_2} shows that the SiO emission peak always occurs in the cool
and dense postshock region: $T_{\rm n} \approx 50$~K (Fig.~\ref{phys_cond_2}a) and
a density of 10--40 times the preshock value, close to maximum
compression (Fig.~\ref{phys_cond_2}b). The SiO fractional abundance, $n({\rm SiO})/n_{\rm H}$, is
also approximately equal to its maximum value in the shock wave (compare
Fig.~\ref{phys_cond_2}c with Fig.~\ref{sio_1}). The trend to lower SiO abundance at higher $n_{\rm H}$, noted in
Section~\ref{sub:grid}, is clearly visible here. Finally, the ``LVG parameter'', $n({\rm SiO})/({\rm d}v_z/{\rm d}z)$, lies typically in the range
$10^{14}-10^{16}$ cm$^{-2}$ km$^{-1}$ s, implying that the 5--4 line is optically thick at its peak for most models of our grid.
In Section~\ref{sub:sio_5}, we compare these physical parameters to values 
inferred previously from LVG analyses of observations, assuming a uniform slab which  fills the beam. 

\subsection{Line profiles and peak line temperatures}
\label{sub:sio_2}

\begin{figure}
\includegraphics[height=20cm]{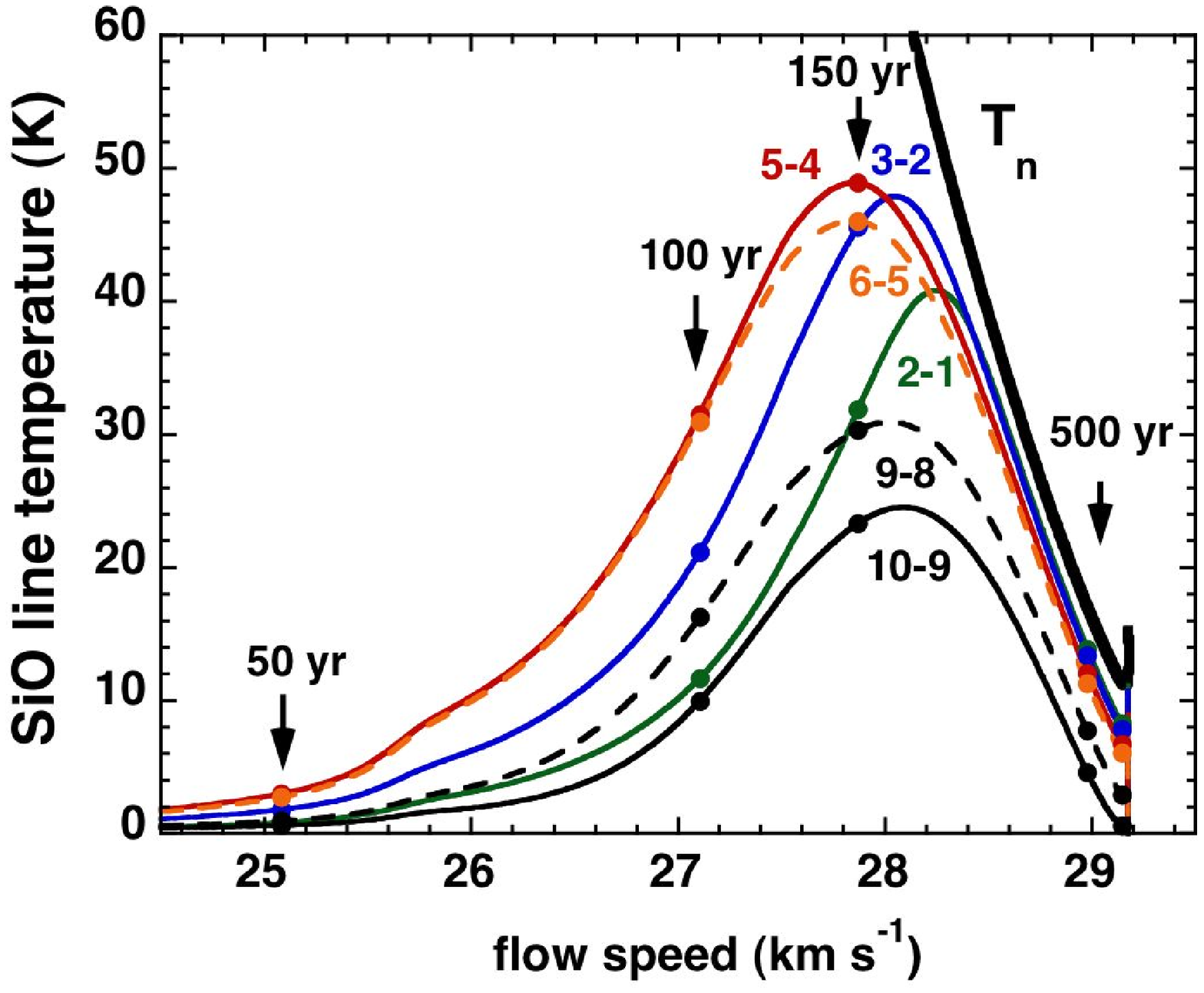}
\caption{The velocity profiles of transitions from rotational levels $j \rightarrow j-1$ of SiO, computed for our reference model, viewed face--on. Only those lines detectable from the
ground are shown. The model parameters are $n_{\rm H} = 10^{5}$~cm$^{-3}$, $B = 200$~$\mu $G, and $v_{\rm s} = 30$~km~s$^{-1}$. The flow speed is in the reference frame of the preshock gas. The neutral temperature profile, $T_{\rm n}$, is shown also, as are indicative values of the flow time of the neutral fluid. (See also Fig.~\ref{O2_8} of Appendix~\ref{O_2}.)}
\label{profiles}
\end{figure}

In Fig.~\ref{profiles}, we compare the intensity profiles of various rotational lines, as
functions of the flow speed of the neutral fluid, expressed in the frame of the preshock gas, for our reference model: $n_{\rm H} = 10^{5}$~cm$^{-3}$, $v_{\rm s} = 30$~km~s$^{-1}$, and $b = 0.63$. The profiles are seen to be narrow (widths of 1--2~km~s$^{-1}$), with 
similar shapes and peaking within 2~km~s$^{-1}$ of $v_{\rm s}$, as expected for compressed material at the rear
of the shock wave.\footnote{The maximum compression of the postshock relative to the preshock gas, $\sqrt{2}v_{\rm s}/v_{\rm A}$, where $v_{\rm s}$ is the shock speed and $v_{\rm A}$ is the Alfv\'{e}n speed in the preshock gas, occurs when the magnetic pressure in the postshock gas is equal to the initial ram pressure. It follows that the flow speed in the postshock gas cannot fall below $v_{\rm min} = v_{\rm A}/\sqrt{2} = 1.3b$~km~s$^{-1}$ {\it in the shock frame}, where $b$ is the scaling parameter of the magnetic field, defined in Section~\ref{sub:grid}.} Note that the transition 2--1 peaks further into the cooling flow, because the lower $j$--levels are
repopulated from the higher levels as the temperature falls.

The general shape and centroid velocities are globally similar to those found
by Sch97 (their fig.~3b where the profiles
were plotted {\it in the shock frame}), although the differences
between the various lines are less significant in the present calculations. Also, the emission wing from the fast precursor, at the start of the shock wave, is weaker in the current models, owing to the delay in SiO formation (see Section~\ref{sub:comparison}), making our line profiles
narrower than in Sch97. Note that including local thermal
broadening in our profile calculations would not
significantly change our predicted SiO line width of 1--2~km~s${-1}$,
because the SiO emission peaks at low temperatures, $T_n \lesssim 100$~K, where the
Doppler width is $\sqrt{kT/44 m_{\rm H}} \le 0.1$~km~s$^{-1}$.

In the left column of Figure~\ref{sio_int_1}, we show the predicted peak
temperature of the SiO 5--4 line for the grid of models considered in
Section~\ref{sub:grid}, as well as the variation with $j_{\rm up}$ of the peak brightness temperatures of various lines, relative to that of 5--4. The {\it
relative} intensities have the advantage of being independent of the
beam filling factor, and thus they are comparable directly to observations,
without prior knowledge of the source size.

The relative peak intensities are within 20\% of unity for $j_{\rm up} \le 7$ over a broad range of model parameters ($v_{\rm s} \ge
30$~km~s$^{-1}$ and $n_{\rm H} < 10^6$~cm$^{-3}$), owing to the large opacity
and near--LTE excitation conditions. For larger values of $j_{\rm up}$, the relative
intensities are more dependent on the shock speed and 
$\approx 1$ only when the limiting speed is approached. The absolute peak brightness temperature in the 
5--4 line is typically 10--50~K for $v_{\rm s} \ge 30$~km~s$^{-1}$, similar to the kinetic temperature in the emission region, but drops sharply at lower shock speeds, for which the SiO abundance (and opacity) is small. The broken curves in Figs.~\ref{sio_int_1}d and h are the results obtained assuming that the initial abundance of O$_2$ ice is negligible, i.e. the second of the two scenarios described in Section~\ref{sub:comparison}.

\subsection{Integrated line intensities}
\label{sub:sio_3}

\begin{figure}
\includegraphics[height=20cm]{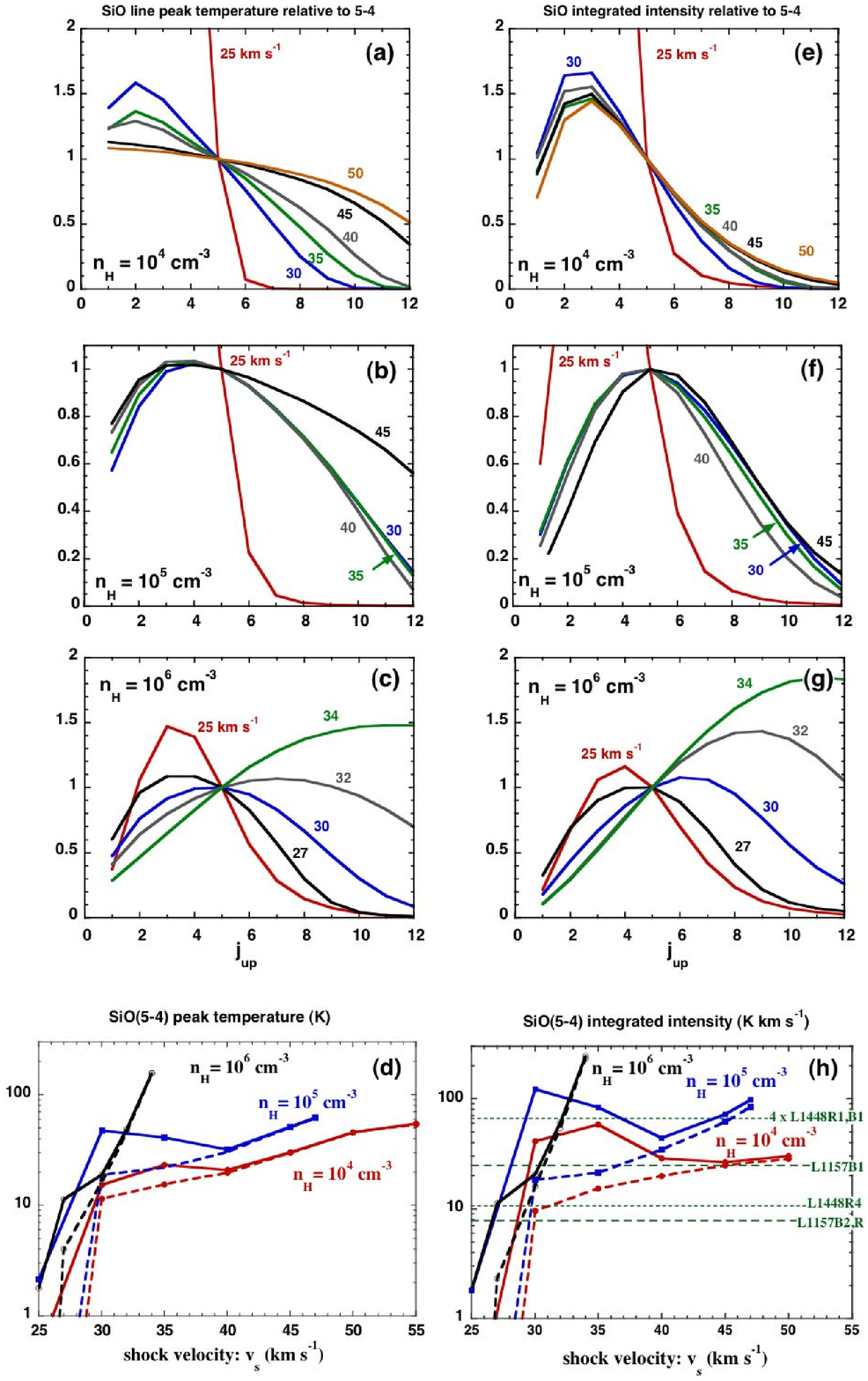}
\caption{(a--c) The peak line temperatures, $T_{\rm peak}$, of the rotational emission lines of SiO, relative to the 5--4 line, as functions of the rotational quantum number of the upper level of the transition, $j_{\rm up}$, for the grid of models in Section~\ref{sub:grid}. The value of density, $n_{\rm H}$, of the preshock gas is indicated in each panel. (d) The {\it absolute} peak
brightness temperature of the 5--4 line, $T_{\rm peak}(5-4)$, as a function of shock speed, $v_{\rm s}$, for all three values of the preshock gas density, $n_{\rm H}$. Shock speeds in excess of 34~km~s$^{-1}$ are absent when $n_{\rm H} = 10^{6}$~cm$^{-3}$, as they give rise to a J-type discontinuity, and the shock wave is no longer C-type; see Section~\ref{sub:grid}. The right--hand panels show the corresponding values of the integrated line intensities, $T{\rm d}V$. The values of $T{\rm d}V(5-4)$ observed in L1157 and L1448 are indicated. In panels (d) and (h), the broken curves show the results obtained assuming that the initial abundance of O$_2$ ice is negligible, i.e. the second of the two scenarios described in Section~\ref{sub:comparison}.}
\label{sio_int_1}
\end{figure}

In the right column of Fig.~\ref{sio_int_1} are presented the integrated intensities (denoted $T{\rm d}V$) of the rotational emission lines of SiO, relative to the 5--4 line, computed for the grid of models considered in Section~\ref{sub:grid}. There are significant differences between the relative integrated and peak ($T_{\rm peak}$) line temperatures (right and left columns, respectively, of Fig.~\ref{sio_int_1}), owing to systematic variations in linewidth
with $j_{\rm up}$, i.e. in the extent of the region where the line is
significantly excited. The relative integrated intensities of lines with
$j_{\rm up} \ge 7$ remain the most sensitive to the shock speed. As may be seen in Sch97, the maximum value of $T{\rm d}V$ occurs at higher values of the rotational quantum number, $j_{\rm up}$, for higher $n_{\rm H}$.
Although the shock temperature varies only weakly with $n_{\rm H}$ (see
Fig.~\ref{sio_1}), a higher density enhances the rates of collisional
excitation of the high--$j$ levels, for any given value of the shock speed, $v_{\rm s}$. In Section~\ref{observations}, we explore the usefulness of this effect for
constraining the preshock density, based on a comparison with actual
observations.

The variation of the {\it absolute} integrated intensity of the 5--4 rotational emission line with the shock parameters is shown also in Fig.~\ref{sio_int_1}h. The bump in $T{\rm d}V$ of the 5--4 transition, for $30 \leq v_{\rm s}
\leq 40$~km~s$^{-1}$ and $10^4 \leq n_{\rm H}
\leq 10^5$~cm$^{-3}$ is caused by incomplete O$_2$
destruction in the shock wave, resulting in more rapid SiO formation and warmer
emission zones (cf. Fig.~\ref{phys_cond_2}). As the broken
curves in Fig.~\ref{sio_int_1}d and h show, this ``bump'' is absent in our second scenario, where O$_2$ is never abundant in the gas phase. At higher shock speeds, the
results from the two scenarios become identical, as OH dominates the
oxidation of Si in both cases.

\subsection{Influence of viewing angle}
\label{sub:sio_4}

\begin{figure}
\includegraphics[height=20cm]{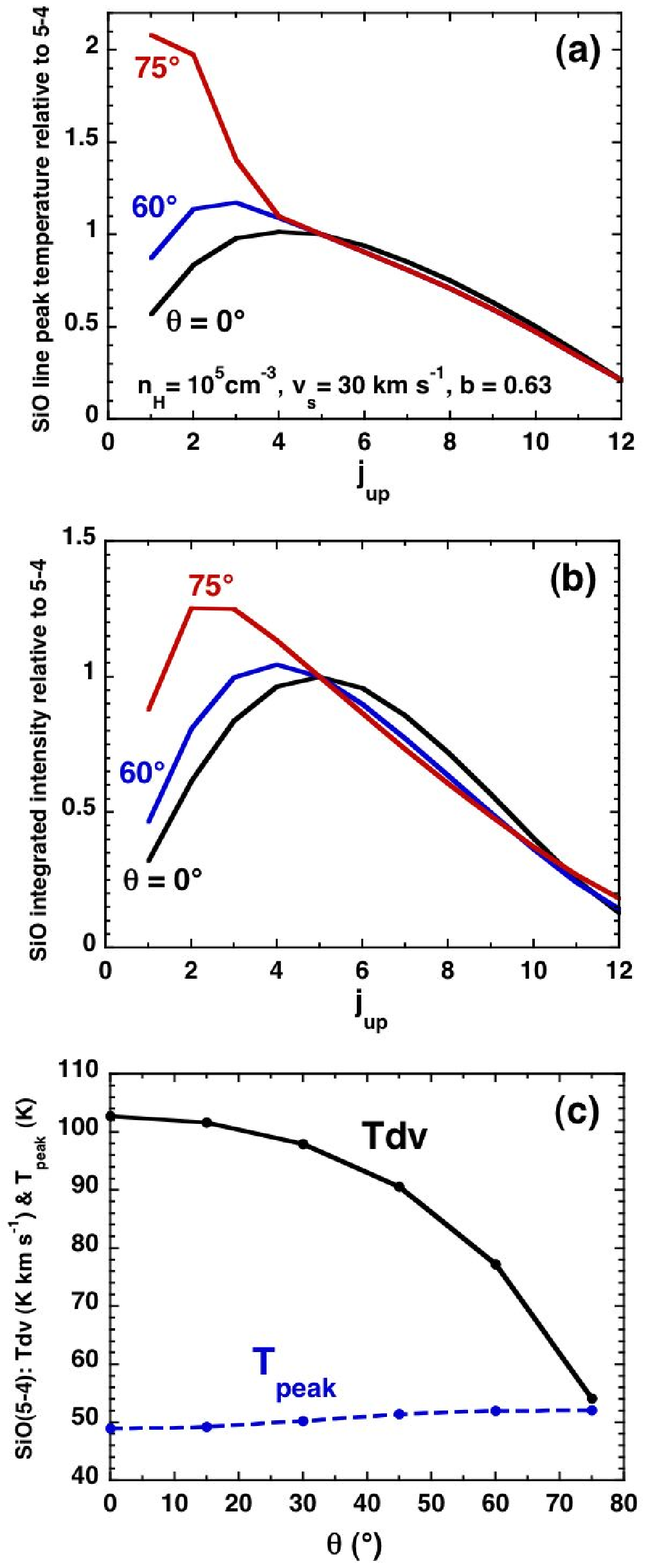}
\caption{Effect of the inclination angle, $\theta $, on (a) the peak, and (b) the integrated intensities of the rotational emission lines of SiO, relative to the 5--4 line; (c) the integrated and peak intensities of the 5--4 line.}
\label{inc_angle}
\end{figure}

Statistically, there is a low probability that a planar shock should happen to be viewed face--on.  Accordingly, we have explored the effects on the SiO rotational line intensities of varying the viewing angle, for the case of our reference model, using the formula~\ref{eq:incl}, derived in Appendix~\ref{appendix}. 

The variations of $T_{\rm peak}(5-4)$ and $T{\rm d}V(5-4)$ with viewing angle, $\theta $, 
are plotted in the bottom panel of Fig.~\ref{inc_angle}. The peak intensity is
almost unaffected by the inclination, as the line is already optically
thick for a face--on view (see Equ.~\ref{eq:incl}). On the other hand, the velocity
projection reduces the line width, and the integrated intensity, $T{\rm d}V$,
declines steadily with increasing viewing angle -- by up to a
factor of 2 at 75$^o$. However, such a variation would be difficult to deduce from
observations, given the typical uncertainties in beam filling factors.

Panels (a) and (b) of Fig.~\ref{inc_angle} illustrate the changes in the (relative to 5--4) peak and integrated SiO line temperatures.  Significant changes, compared with viewing face--on, are seen only for
inclinations greater than 60$^o$ from the normal, and they affect only the
optically thin lines $j_{\rm up} \le 3$ and $j_{\rm up} \ge 12$. The main change
in the curves for the relative integrated line temperatures (panel (b) of Fig.~\ref{inc_angle}) is that the maximum occurs at lower $j_{\rm up}$ as $\theta $ increases; this effect could
be easily confused with a face--on shock of slightly lower preshock
density (cf. Fig.~\ref{sio_int_1}). The relative peak temperature (panel (a)) is even more strongly
modified, with an upward turn of the curve at $j_{\rm up} \le 4$.  The latter
characteristic appears to be the only unambiguous signature of a
viewing angle $> 60^o$ in the context of our one--dimensional models.

\subsection{Influence of the transverse magnetic field strength}
\label{sub:sio_5}

\begin{figure}
\includegraphics[height=20cm]{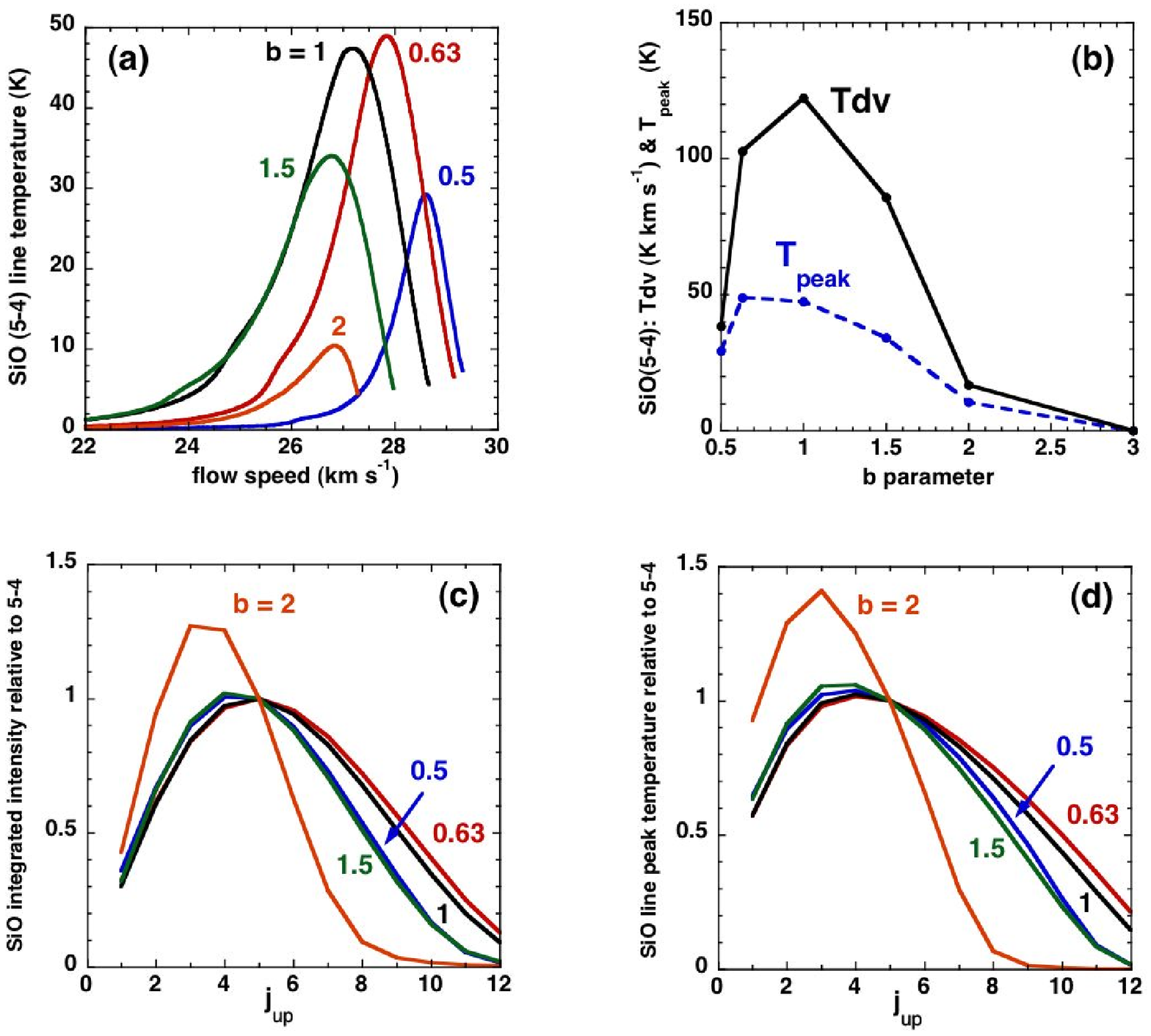}
\caption{(a) The SiO 5--4 line temperature, as a function of the flow speed of the neutral fluid, in the reference frame of the preshock gas, for the specified values of the magnetic field parameter, $b$; (b) the peak and integrated intensities of the 5--4 line, as functions of $b$; (c) the integrated and (d) the peak intensities of rotational emission lines $j_{\rm up} \rightarrow j_{\rm up}-1$ of SiO, relative to the 5--4 transition, for the specified values of $b$. All calculations for $v_{\rm s} = 30$~km~s$^{-1}$ and $n_{\rm H} = 10^{5}$~cm$^{-3}$.}
\label{sio_rel_b}
\end{figure}

In Section~\ref{sub:magnetic}, we have seen that the efficiency of sputtering Si from
grains decreases monotonically with increasing magnetic
field strength. The effect of varying the scaling parameter, $b$, on the emergent SiO line intensities is shown in Fig.~\ref{sio_rel_b}, for our reference model.

In panels (a) and (b) of Fig.~\ref{sio_rel_b} are plotted the predicted line profiles, peak temperatures, and integrated intensities of the SiO 5--4 line, for
various values of $b$. It may be seen that the maximum intensity is reached for
intermediate values of $0.63 \lesssim b \lesssim 1$. At smaller $b$, SiO is less
abundant: the shock wave is narrower and hotter, and so O$_2$ is more readily 
destroyed by H, resulting in incomplete oxidation
of Si into SiO. 
At larger $b$, the SiO emission decreases owing to less efficient
sputtering of Si (see Fig.~\ref{Bfield}b) at the lower ion--neutral drift speeds. In particular, the predicted intensity drops from $T_{\rm peak} = 10$~K at $b = 2$ to practically zero at $b = 3$.  

Panels (c) and (d) of Fig.~\ref{sio_rel_b} show the relative peak and integrated
line temperatures, as functions of $j_{\rm up}$, for various values of $b \le 2$
(curves for $b > 2$ are not shown as they lead to negligible SiO
emission).  It may be seen that the values $0.63 \lesssim b \lesssim 1$, which give rise to 
the strongest 5--4 emission, also yield the highest relative
intensities of lines from $j_{\rm up} \ge 7$.  For example, the intensity of the 
11--10 line, relative to 5--4, is 3 to 4 times larger than when $b = 0.5$ or
$b = 1.5$.  Comparison with Fig.~\ref{sio_int_1} suggests that this dependence on
$b$ may be difficult to distinguish observationally from variations in
shock speed. A less ambiguous indication of the value of $b$ might be obtained from
the width of the cooling zone, which increases as $b^2$, from $10^{15}$~cm for $b = 0.4$ to $2\times 10^{16}$~cm for $b = 2$ and for the parameters of our reference model (see Fig.~\ref{Bfield}a).

\section{Comparisons with observations}
\label{observations}

\subsection{Rotational line profiles}

As Sch97 first noted, the generic SiO line profile
predicted by steady planar C-type shock waves, with a peak at high velocity
(in the postshock gas) and a tail at lower velocity (in the accelerating precursor),
is reminiscent of the SiO line profiles in the L1448 molecular
jet (Bachiller et al. 1991). Similarly, we note
that the reversed shape of SiO profiles in the L1157 bowshocks, with
a peak at low velocity and a high velocity tail (Zhang et al. 1995),
could arise if the postshock gas is stationary in the cloud frame, i.e. if
one observes the reverse shock, in which the jet is being
decelerated. However, in either case, the SiO profiles predicted by 
our models remain narrower than those observed, with
widths of 0.5--2~km~s$^{-1}$, as compared to the observed widths of
5--20~km~s$^{-1}$ in 3\arcsec\--10\arcsec\ beams.

Broader line profiles from steady C-type shocks could arise if Si was
sputtered not only from grain cores, as assumed here, but also from
SiO--containing grain mantles, with lower binding energy. Then, the SiO abundance would be much enhanced at intermediate velocities, in the precursor
(see Sch97). However, owing to the steep temperature decline across the shock wave,
this situation results in large variations of the line
widths and velocity centroids with the emitting rotational level, $j_{\rm up}$ (cf. fig.~5 of Sch97). In fact, the observed profiles are very similar from line to
line, with the (8--7)/(2--1) intensity ratio showing only modest variations with
velocity (see, for example, fig.~9 of Nisini et al. 2007). These observations suggest that the broad SiO lines are not
attributable entirely to intrinsic velocity gradients through a single,
planar C-type shock wave. There may be several shock--cooling zones inside the beam, each with a narrow intrinsic profile,
which appear spread out in radial velocity owing to a range of
inclination angles or propagation speeds, in the observer's frame. This conclusion is supported by interferometric observations of L1157 and L1448 
(Guilloteau et al. 1992; Gueth et al. 1998; Benedettini et al. 2007), which reveal
systematic velocity gradients across the SiO emitting knots (reminiscent, in some cases, of a bowshock geometry; Dutrey et al. 1997) down to 2\arcsec\--3\arcsec\ 
resolution; at the distances of L1157 and L1448, the angular dimensions of the SiO emitting regions in Fig.~\ref{profiles}, for example, are a few tenths of an arcsec. Such complex two--dimensional modelling lies outside
the scope of the present paper. However, we argue in Section~\ref{line_intensities} that we may still perform a meaningful comparison of our predicted 
SiO line intensities with observations of knots in outflows, without
reproducing in detail the line profiles, provided that the shock conditions do not
vary too much across the beam.

\subsection{Narrow SiO lines near ambient velocity}

In addition to the typically broad SiO line profiles mentioned
above, some outflow regions such as NGC1333 and L1448 exhibit
extremely narrow SiO emission lines, with $\Delta v \approx 0.5$~km~s$^{-1}$, near rest velocity (Lefloch et al. 1998; Codella et al. 1999;
Jim\'enez-Serra et al. 2004, 2005). The corresponding SiO abundance of 
$10^{-11}$--$10^{-10}$ is two to three orders of magnitude smaller than in the
broad SiO components (Codella et al. 1999; Jim\'enez-Serra et
al. 2005). Jim\'enez-Serra et al. proposed that this feature in L1448 
traces a magnetic precursor, where neutral gas is just
beginning to accelerate and grain species are starting to be released into the
gas phase. However, as we now explain, detailed multifluid shock models do not support this interpretation.

Our SiO line profile calculations show that emission from a
magnetic precursor does not give rise to a narrow feature near
the speed of the preshock gas. The line intensity increases as the
neutral fluid is accelerated, heated, and enriched in SiO -- 
by orders of magnitude by the time that grain--sputtering is
complete. Indeed, this deduction could have been made already, on the basis of figs.~3 and 5 of Sch97, which cover the entire velocity range relevant to predicting the SiO line profiles. Truncation of the precursor when the neutral fluid has been accelerated to only $v_{\rm n} = 0.5$~km~s$^{-1}$ would imply a very finely--tuned shock age, a circumstance which appears to us to be improbable.
Furthermore, an ion--neutral drift speed of at least 5~km~s$^{-1}$ is needed
to start releasing species from grain mantles, where binding energies
are a few tenths of an eV (Flower and Pineau des For\^ets 1994), and of
at least 20~km~s$^{-1}$ to start sputtering grain cores (May et
al. 2000). However, the H$^{13}$CO$^+$ line does not show evidence of this
predicted acceleration: its emission peak is shifted by only $+0.5$~km~s$^{-1}$
from the velocity of the ambient gas, like the narrow SiO feature (Jim\'{e}nez-Serra et al. 2004).

We believe that a more likely explanation of the narrow SiO feature in
L1448 is that it traces Si--enriched postshock material that has been
decelerated by and mixed with the ambient gas, as
proposed originally by Lefloch et al. (1998) and Codella et al. (1999) 
in connection with other regions. Given a shock speed $v_{\rm s} \leq 30$~km~s$^{-1}$ and the high ambient density characteristic of Class 0 protostellar envelopes,
deceleration could be achieved readily within the L1448 flow age of approximately 
3500 yr. The low SiO fractional abundance would then be a consequence of mixing with
SiO--poor ambient gas. Alternatively, the narrow feature might arise 
in a reverse C-type shock, where outflow material at $v < 20$~km~s$^{-1}$
is brought almost to rest by the much denser ambient medium, and the shock
speed is too low to produce abundant SiO. Both interpretations are consistent
with NH$_3$ observations of dense gas in the envelope of the L1448
protostar, with radial velocity and spatial extent similar to that of
the narrow SiO feature and signs of heating near the path of the fast
L1448 jet (Curiel et al. 1999).

\subsection{SiO line intensities}
\label{line_intensities}

If our explanation of SiO profile broadening is correct, one
could in principle recover the parameters of {\it each} individual
emission zone in the beam by analysing the relative intensity ratios
{\it as functions of velocity}. Unfortunately, such data are currently
quite noisy and not yet available for a wide range of values of 
$j_{\rm up}$. Furthermore, knowledge of the beam--filling factor as a function
of velocity would be necessary to obtain absolute intensities and remove
ambiguity in the shock parameters; but this would require sub-arcsecond angular resolution, which is not yet available.
Nevertheless, one may still derive some approximate {\it
beam--averaged} shock properties, if all of the shock components have similar
excitation conditions, as is suggested by single--dish data, which show the line
ratios to be insensitive to velocity, $v$. In this case, the observed
profile will be simply a convolution of the individual, narrow shock
profiles with the (unknown) filling--factor, $\phi(v)$. The observed 
absolute $T{\rm d}V$ is simply that for a single shock, multiplied by the total
beam filling factor of the SiO--emitting region in the beam, $f = \int{\phi(v) {\rm d}v}$, as inferred from its overall size in single--dish maps. The values of $T{\rm d}V$ for different $j_{\rm up}$, relative to the 5--4 transition, remain unchanged compared to a single shock, because $f$ cancels out in the ratios, thereby enabling direct comparison with our models. In the following, we assume that this situation prevails.

\begin{figure}
\includegraphics[height=20cm]{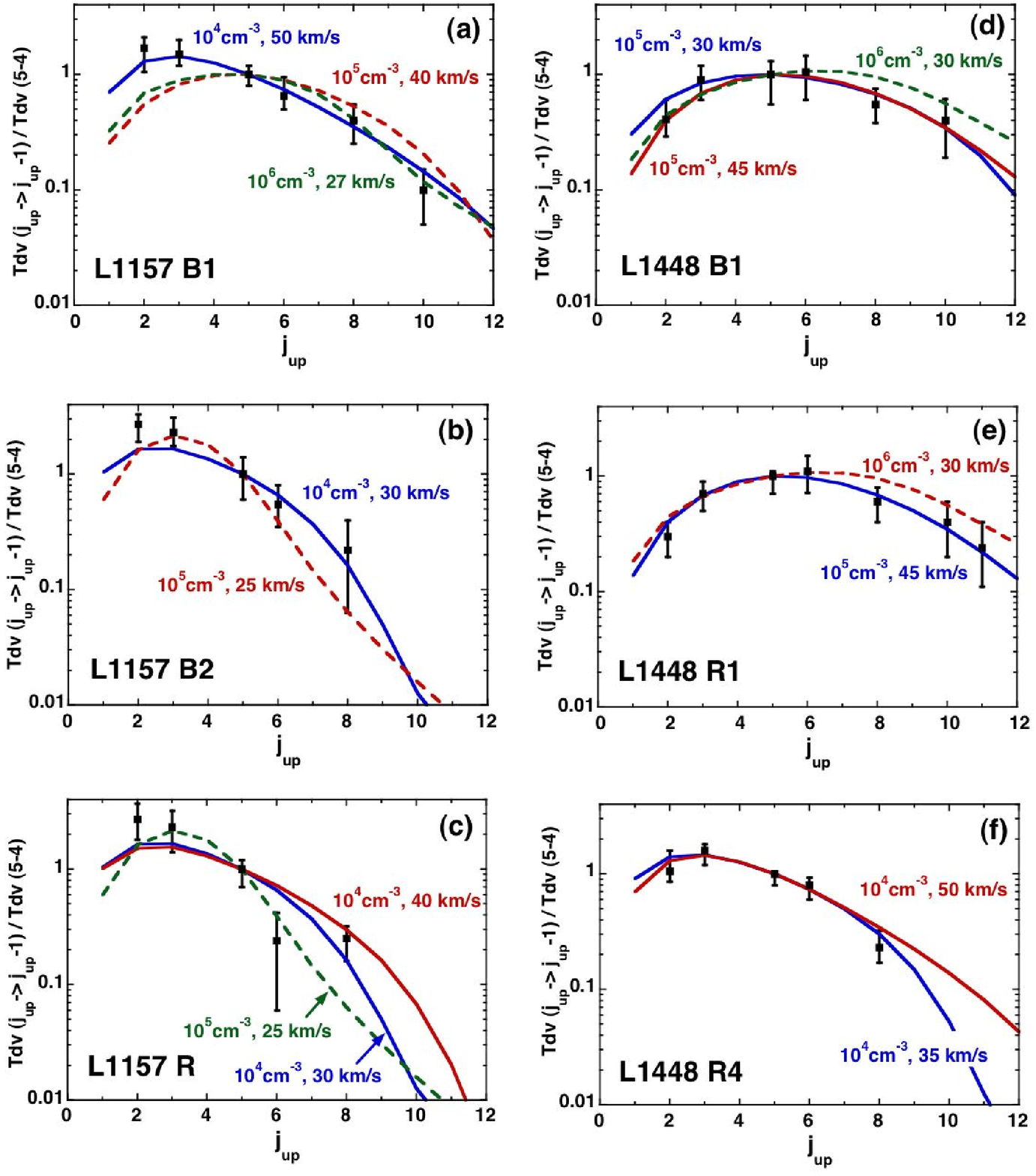}
\caption{The relative intensities of the rotational emission lines of SiO observed in the outflow sources L1157 and L1448 (Nisini et al. 2007: points with error bars) and predicted by the C-type shock models (curves) with the parameters ($n_{\rm H}$, $v_{\rm s}$) indicated; see text, Section~\ref{observations}. The data points which are plotted include a correction for differing beam sizes, which is significant for low--$j$ lines. Full curves denote the best--fitting models of our grid. Broken curves show ``near--miss'' models with a different value of $n_{\rm H}$, which either yield a worse fit to the data points or do not reproduce the absolute $T{\rm d}V$ of the 5--4 line. (See also Fig.~\ref{O2_12} of Appendix~\ref{O_2}.)}
\label{sio_rel}
\end{figure}

By way of illustration of the applicability of the shock models, we
show in Fig.~\ref{sio_rel} the relative integrated intensities of the
rotational transitions of SiO observed in the outflow sources L1157
and L1448 (Nisini et al. 2007) and predicted by the grid of models,
whose parameters are specified; in all cases, the magnetic field
scaling parameter $b = 1$. As noted in Section~\ref{sub:sio_5}, this
value of $b$ yields the largest relative intensities of the high--$j$
lines and hence will yield a {\it lower limit} to the shock speed
required to reproduce the observations (except at $n_{\rm H} =
10^5$~cm$^{-3}$, where high--$j$ excitation is a non-monotonic
function of $v_{\rm s}$; see Fig.~\ref{sio_int_1}f). We assume also
that the shock wave is viewed face--on; if the true inclination
exceeds $60^o$, this assumption results in the preshock density being
slightly underestimated (see Section~\ref{sub:sio_4}). The models
shown as the full curves are those which provide the best fits to the
observations. In order to illustrate how well the shock parameters are
constrained, we plot as dashed curves ``near--miss'' models that fit
most of the data points, or fit all points but do not reproduce the
absolute intensity of the 5--4 line (see below).

Fig.~\ref{sio_rel} demonstrates that steady--state C-type shocks 
with Si--sputtering from grain cores can
reproduce successfully the relative integrated intensities of SiO
lines in these molecular outflows. Furthermore, Fig.~\ref{sio_int_1}h
shows that the models can reproduce also the {\it absolute} integrated
intensity of SiO 5--4 with the estimated beam filling factors $f
\approx 1$ in L1157 and L1448-R4 
and $f \simeq 1/4$ in L1448 R1 and B1 (cf. Nisini et
al. 2007).  Alternative fits with higher density and lower shock
speeds ($n_{\rm H} = 10^6$~cm$^{-3}$ and $27 \lesssim v_{\rm s}
\lesssim 30$~km~s$^{-1}$; $n_{\rm H} = 10^5$~cm$^{-3}$ and $v_{\rm s}
= 25$~km~s$^{-1}$) underestimate $T{\rm d}V(5-4)$ and are thus ruled
out. The shock speed, $v_{\rm s}$, is constrained to within
15~km~s$^{-1}$ and the preshock density to within a factor of 10, with
one of our grid values of $n_{\rm H}$ yielding a clear best fit in all
cases.

In Appendix~\ref{O_2}, we present, in Fig.~\ref{O2_12}, results equivalent to
those in Fig.~\ref{sio_rel}, but for the secondary grid of models, in which 
oxygen is initially in the form of H$_2$O ice rather than O$_2$ ice. Again, the relative intensities can be well reproduced by
steady--state C-type shocks. The absolute SiO $T{\rm d}V(5-4)$ favour the
low $n_{\rm H}$, high $v_{\rm s}$ cases (cf. the dashed curved in Fig.~\ref{sio_int_1}h).
The best--fit shock parameters and the inferred physical
conditions at the (5--4) line peak are almost unchanged, as
the range of shock speeds is such that oxidation of Si by OH is dominant.

It is instructive to compare the physical parameters at the SiO peak
of our best--fit, steady--state C-type shock models to those
previously inferred from an LVG analysis, assuming a slab of
constant density, temperature, and velocity gradient along the line of
sight (Nisini et al. 2007). From Table 1, it may be seen that the
`slab LVG' approach yields similar values of the density to our shock
models but overestimates by a factor 5--10 the kinetic temperature
and underestimates by several orders of magnitude the Sobolev LVG
opacity parameter. The cool postshock layer emits over a narrow
velocity range and needs to be more optically thick in SiO
to produce the same $T{\rm d}V$ as a hot slab with a large velocity gradient
along the line of sight. On the other hand, similar SiO abundances are deduced using both approaches, to within typically a factor of 3.

We note that the models should be able to simulate also the other spectral observations of the sources, notably the H$_2$ line intensities. In a forthcoming publication, we shall consider in detail the outflow L1157 and make a more comprehensive comparison of its observed spectrum with the predictions of shock models.

\begin{table}

\begin{tabular}{l | l l l l l l | l l l l}
\hline
\multicolumn{6}{c|}{Face--on C--type shock models$^{a,b}$} &
\multicolumn{4}{c}{Slab LVG models$^{a,c}$}\\SiO knot &   $n_{\rm H}^{\rm
init}$ & $v_{\rm s}$  &  $T_{\rm kin}$  & $n_{\rm H}$  &  LVG &  $x_{\rm
SiO}$ & $T_{\rm kin}$ & $n_{\rm H}$  & LVG & $x_{\rm SiO}$ \\
\hline
L1448
B1 &  $1(5)$ & 30,45 & 90--70 &10--30 & 2--4 &  0.5--2 &$>$500 & 8 & 0.1 &
1 \\
L1448 R1 &  $1(5)$ & 45    & 70 &   30   & 4 &  0.5 &  $>$500  &  10 &
0.1 & 1\\
L1448 R4 &  $1(4)$ & 35--50 & 45--55 &1.5--4 &7--100 &3--7 & 200
&   2.5 & 0.03 & 0.3\\
\hline
L1157 B1 & $1(4)$  & 50 & 55 & 4 & 100 & 3 &
150--300 & 3 & 0.08 & 0.8 \\
L1157 B2 & $1(4)$  & 30	 &35 & 1.5 & 4 & 2
& 200--300 &2 & 0.05 & \\
L1157 R  & $1(4)$  & 30--40 &35--30 &  1.5--2 &
4--10 & 2--3 & 50--100 &1--5 & 0.02 & 0.6\\
\hline

\end{tabular}

$^a$Units: $n_{\rm H}^{\rm init}$ is the preshock density,
in cm$^{-3}$; $v_{\rm s}$ the shock speed, in km~s$^{-1}$; $T_{\rm kin}$ is the
local gas kinetic temperature, in K; $n_{\rm H}$ is the local gas density, {\bf
in $10^5$~cm$^{-3}$; ``LVG'' is the local LVG parameter, $n$(SiO)/$(dv_{\rm n}/dz)$,
in $10^{14}$~cm$^{-2}$~km$^{-1}$~s; and $x_{\rm SiO}$ is the local
fractional abundance, in units of $10^{-7}$}.\\ $^b$Best grid model
from Fig.~\ref{sio_rel} and physical parameters at the SiO 5--4 line
peak from Fig.~\ref{phys_cond_2}.\\
$^c$Values taken from Tables~4 and 5 of Nisini
et al. (2007). The LVG parameter is given by $N$(SiO)/$\Delta V$,
with $\Delta V = 10$~km~s$^{-1}$.\\
\caption{Properties of SiO--emission
regions deduced from face--on C-type shock models or homogeneous--slab LVG
models.}

\end{table}

\section{Concluding remarks}
\label{sec:conclusion}

We have considered the structure of C-type shock waves propagating into molecular gas containing amorphous carbon and silicate grains; olivine (MgFeSiO$_4$) was chosen as the representative silicate--grain material. We find that the degree of sputtering of silicon from the grains is smaller, by about an order of magnitude, than was predicted by the calculations of Sch97, owing partly to higher sputtering thresholds and lower sputtering yields but principally to the reduced width of the shock wave in the present calculations. The reduction in the shock width is a consequence of a more accurate treatment of the coupling between the neutral fluid and the charged grains in the current model. 

A grid of C-type shock models has been computed, for values of the preshock gas density and the shock speed which are believed to span the ranges of these parameters in molecular outflows; two scenarios were considered regarding the initial distribution of oxygen in the gas and solid phases. The maximum speed of a C-type shock is limited by the inertia of the (charged) grains in the preshock gas and by the collisional dissociation of molecular hydrogen within the shock wave, which can lead to a J-type discontinuity (Le Bourlot et al. 2002, Flower \& Pineau des For\^{e}ts 2003). We find that significant sputtering of the grain cores occurs only for shock speeds $v_{\rm s} \ge 25$~km~s$^{-1}$ and moderate magnetic field strengths close to equipartition with the cloud kinetic energy (magnetic field parameter $0.5 \le b \le 2$). However, we note that the sputtering threshold energy is determined principally by the so--called ``displacement energy'', $E_{\rm D}$, of the material, whose value for the silicates of relevance here remains uncertain (May et al. 2000). The sputtering yields increase rapidly from threshold, and a reduction in the threshold energy would enable significant sputtering to occur at lower shock speeds or higher magnetic field
strengths.

We find that, in the absence of silicon--containing grain mantles, SiO line emission in steady--state planar C-type shock waves arises
predominantly from cool postshock gas, close to maximum compression,
with negligible emission from the precursor. Except at the lowest
shock speeds, the SiO emission is optically thick and close to LTE for
$4 \le j_{\rm up} \le 7$.  The relative line intensities, as functions of
$j_{\rm up}$, together with the absolute 5--4 line intensity, provide good
diagnostics of the shock parameters, $n_{\rm H}$ and $v_{\rm s}$. The influence of the viewing angle and transverse magnetic field strength is found to be relatively
minor over the typical ranges of their values. 

Our shock models provide good fits to both the relative and absolute SiO
intensities in the molecular outflows L1157 and L1448; the SiO fractional 
abundance is deduced to be in the range $4\times 10^{-8} \lesssim n({\rm SiO})/n_{\rm H} \lesssim 3\times 10^{-7}$. The emission regions of the shock wave are much colder,  more optically thick, and have 10 to 100 times greater SiO column density than
estimated previously from optically thin LVG slab models (Nisini et
al. 2007). Our results are in line with a recent analysis of interferometric
maps of the HH212 jet (Cabrit et al. 2007), demonstrating that the SiO emission
is optically thick and close to LTE, with an intrinsic peak brightness
temperature of approximately 50~K. 
On the other hand, the line profiles predicted by our planar C-type shocks 
are typically 10 times narrower than observed in L1157 and L1448, suggesting that the 
single--dish beam includes shocks with various inclinations and speeds, and/or mixing layers. 
Detailed modelling of the line ratios as functions of velocity,
with a careful correction for differing beam widths, would be needed to 
clarify this issue.

Whilst the grid of models presented here is intended to provide a guide to interpreting observations of outflow sources, it should be recalled that the dynamical timescales which characterize these regions are often too short to enable C-type shock waves to attain their steady--state structure (Chi\`{e}ze et al. 1998, Lesaffre et al. 2004). The presence of an embedded J-type discontinuity has then to be considered when modelling specific sources. Studies of the outflow source in Orion (Le Bourlot et al. 2002), of jets associated with low--mass star formation (Giannini et al. 2004, 2006; McCoey et al. 2004), and of the supernova remnant IC 443 (Cesarsky et al. 1999) have shown that the rovibrational line spectrum of H$_2$ can be reproduced successfully by hybrid shock waves, but not by pure C- or J-type shocks. The influence of the discontinuity on the C-component (magnetic precursor), and hence on the formation of SiO and its emission line spectrum, becomes important for ages smaller than the flow time to maximum compression.

The existence of Si--containing
grain mantles is another circumstance that would significantly modify the intensities  of SiO lines and their profiles. The fractional abundance of SiO
in the warm shock precursor would be enhanced, compared with the models considered here, in which Si is sputtered exclusively from olivine grain cores. The effects
of non--steady C-type shocks and Si--containing mantles will be considered in a
forthcoming paper.

\begin{acknowledgements}
Antoine Gusdorf and the University of Durham acknowledge the support of the European Commission under the  Marie Curie Research Training Network ``The Molecular Universe'' MRTN-CT-2004-512302. We thank Brunella Nisini for helpful correspondence relating to the SiO observations of \cite{nisini07}. We thank also Paul Goldsmith and Laurent Pagani for information regarding SWAS and Odin observations of O$_2$.
\end{acknowledgements}

\appendix

\section{SiO radiative transfer}
\label{appendix}

\subsection{Photon escape probabilities}

The SiO rotational level populations and excitation temperatures in
our planar shock models are calculated by means of a large velocity
gradient (LVG) method. This approximate treatment assumes that, owing to the macroscopic velocity field and
resulting Doppler shifts, emitted line photons are either re-absorbed locally
or escape to infinity. The escape probability of a line photon in a
given direction $\vec{\hat{s}}$ is then (see Surdej (1977) for a pedagogical derivation):
\begin{equation}
\beta_s = \frac{1 - {\rm e}^{-\tau_s}}{\tau_s}
\end{equation}
where the `LVG optical depth' $\tau_s$ is defined as
\begin{equation}
\label{eq:tau}
\tau_s = \frac{hc}{4 \pi} \frac{n_{\rm l}}{\partial (\vec{v} .\vec{\hat{s}})
/\partial{s}} B_{\rm lu} \left( 1 - \frac{g_{\rm l}n_{\rm u}}{g_{\rm u}n_{\rm l}} \right
),
\end{equation}
with ${\partial (\vec{v} .\vec{\hat{s}})
/\partial{s}}$ the radial velocity gradient along direction $\vec{\hat{s}}$,
$n_l$ and $n_u$ the number density of molecules in the lower and
upper levels of the transition, respectively, $g_{\rm l}$ and $g_{\rm u}$ the
corresponding statistical weights, and $B_{\rm lu}$ the Einstein
coefficient for stimulated absorption.  The mean intensity of the
radiation field at the local frequency $\nu$ of the transition, averaged
over all angles, may then be expressed in terms of local quantities
only as
\begin{equation}
\label{eq:jnu}
\bar{J_\nu} = S_\nu(1-\bar\beta)+I_{\rm c}\bar\beta,
\end{equation}
where $\bar\beta = \int{\beta_s d\Omega}/4\pi$ is the escape
probability averaged over all solid angles; $I_{\rm c}$ is the mean
intensity of the continuum radiation field, taken to be a blackbody
(Planck) function $B_\nu(T)$ at the temperature of the cosmic
background, $T_{\rm bg} = 2.73$~K; and $S_\nu = B_\nu(\mTex)$ is the local
source function, where the excitation temperature \Tex\ of the
transition is defined through $n_{\rm u}/n_{\rm l} \equiv g_{\rm u}/g_{\rm l}
\exp (-h\nu/k_{\rm B} \mTex)$.  Two expressions have been used for the
average escape probability \abeta:

-- that of Neufeld \& Kaufman (1993):
\begin{equation}
\label{eq:betaplane}
\bar\beta_{\rm plane} = \frac{1}{1 + 3\tau_\perp},
\end{equation}
where \ptau\ is the LVG opacity in the $z$-direction, normal to the shock
front. This approximation to \abeta\ is accurate for a
plane--parallel flow, where ${\partial (\vec{v} . \vec{\hat{s}})/ \partial s
} = \mu^2 ({\rm d}v_z/{\rm d}z)$ and $\tau_s = \tau_\perp/\mu^2$ (with the usual notation 
$\mu = \cos\theta = \vec{\hat{s}} . \vec{\hat{z}}$).

\begin{figure}
\includegraphics[height=20cm]{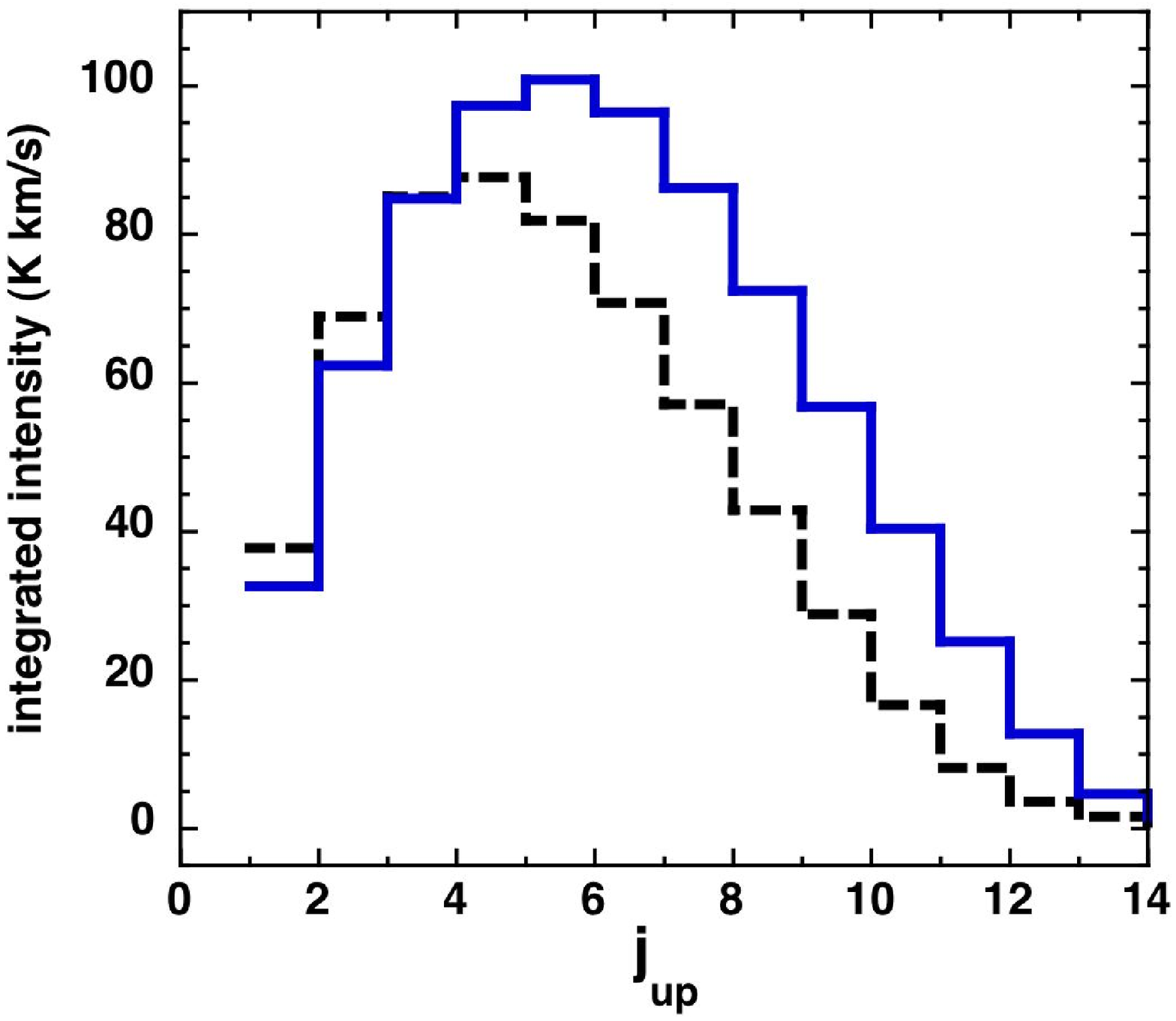}
\caption{The integrated intensities of the rotational transitions of SiO. Full curves correspond to the \cite{neufeld} expression~(\ref{eq:betaplane}) for the escape probability in a plane--parallel flow, dashed curves to the isotropic case,  equation~(\ref{eq:betaiso}). The model parameters are $n_{\rm H} = 10^{5}$~cm$^{-3}$, $B = 200$~$\mu $G, and $v_{\rm s} = 30$~km~s$^{-1}$.}
\label{EP_TAU}
\end{figure}

-- an isotropic approximation: 
\begin{equation}
\label{eq:betaiso}
\bar\beta_{\rm isotropic} = \beta_\perp = \frac{1 - e^{-\tau_\perp}}{\tau_\perp}.
\end{equation}
The first expression was ultimately adopted in the present study for
consistency with our one--dimensional shock geometry, whereas the second was used by
Sch97. Since \abeta\ is always smaller in the
plane--parallel case (owing to the reduced velocity gradients at small
$\mu$), photon trapping is more efficient and the excitation
temperatures are increased compared to the isotropic approximation.
Thus, adopting the expression of Neufeld \& Kaufman (1993) leads to
larger integrated SiO line intensities; this effect is illustrated in
Figure~\ref{EP_TAU} for our reference shock model. It is marginally significant for small and large values of the rotational quantum number but reaches a factor 3 for $j_{\rm up} \approx 12$.

A necessary condition for the local LVG approximation to be valid is
that, over the characteristic distance $L$ where physical conditions
vary, the velocity shift $\Delta v$ arising from the velocity gradient
should be larger than the local line thermal width $v_{\rm th}$. Then, line photons are re-absorbed only within a region of size $< L$,
where physical conditions and SiO excitation are uniform.  This
criterion can be rewritten as
\begin{equation}
\label{eq:lvg}
|{\rm d}v_z/{\rm d}z| > v_{\rm th}/L,
\end{equation}
where $$v_{\rm th} = \sqrt{8k_{\rm B}
T_{\rm n} \over \pi m},$$
$T_{\rm n}$ is the temperature of the neutral fluid, and $m$ is the mass of
the molecule. 

\begin{figure}
\includegraphics[height=20cm]{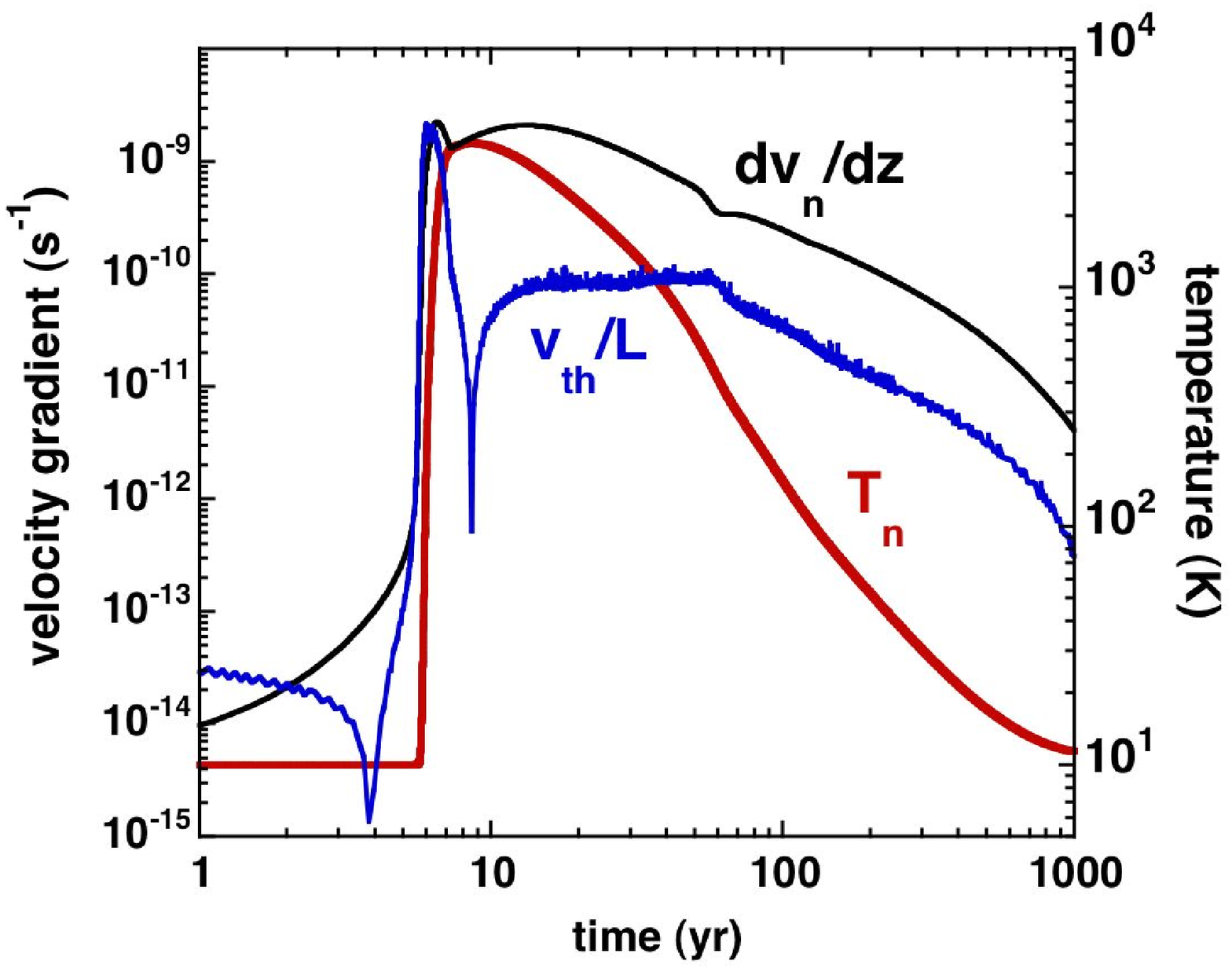}
\caption{Consistency of the LVG method: velocity gradient criterion~(\ref{eq:lvg}). The model parameters are $n_{\rm H} = 10^{5}$~cm$^{-3}$, $B = 200$~$\mu $G, and $v_{\rm s} = 30$~km~s$^{-1}$.}
\label{consistency1}
\end{figure}

In Figure~\ref{consistency1} we compare the left and right hand sides of
(\ref{eq:lvg}) for our reference shock model, using $L \approx
T_{\rm n}/|dT_{\rm n}/dz|$ as a characteristic distance.  The LVG
criterion may be seen to be verified throughout the cooling flow of
the shock wave, where the bulk of SiO emission arises. It is not
verified in the far postshock region, where the computed
velocity gradient tends to zero; but this region makes a negligible contribution
to the line flux, owing to the low escape probabilities.

\subsection{Numerical implementation}
\label{sec:numerics}
In the diatomic SiO molecule, electric dipole transitions take place
only between adjacent rotational levels of the ground vibrational
state ($\Delta j = \pm 1$ with $j$ the rotational quantum number),
while collisionally induced transitions can, in principle, connect any
pair of levels (in practice they become less probable with increasing
level separation).  The evolution of the population density $n_i$ of
level $i$ may thus be written in the following matrix form:
\begin{eqnarray}
\label{eq:matrix}
\frac{dn_{i}}{dt} & = & n_{i+1} (A_{i+1,i} + B_{i+1,i}{\bar{J}_{i,i+1}}) \nonumber \\
		& - & n_{i} (A_{i,i-1} + B_{i,i+1}\bar{J}_{i,i+1} + B_{i,i-1}\bar{J}_{i-1,i}) \nonumber \\
		& + & n_{i-1}B_{i-1,i}\bar{J}_{i-1,i} \nonumber \\
		& + & \sum_{\rm coll} \sum_{j \neq i} n_{\rm  coll}
(n_{j}C_{ji} - n_{i}C_{ij}),
\end{eqnarray}
where $n_{\rm coll}$ is the number density of each collisional partner
(here H$_2$ and He), $C_{ij}$ is the collisional rate coefficient from
level $i$ to level $j$, and $A$ and $B$ denote the Einstein
coefficients for spontaneous and induced radiative transitions.  Two
approaches to solving for the level populations may be
taken:
\begin{itemize}

\item Integrate the set of Equs.~(\ref{eq:matrix}) in parallel with the
dynamical, thermal, and chemical rate equations of the shock wave, as is 
done currently for the H$_2$ molecule (Le Bourlot et al. 2002).

\item Assume local statistical equilibrium, i.e. set the lhs of
Equs.~(\ref{eq:matrix}) to zero, and solve {\it a posteriori} the
resulting algebraic equations by matrix inversion, using the physical
structure provided by the MHD shock code.  Since the radiative terms
${\bar J}$ depend indirectly on the level populations through the
escape probabilities and excitation temperatures (Equ.~(\ref{eq:jnu})),
the procedure must be iterated, updating ${\bar J}$ with
values of \abeta\ and \Tex\ obtained from the previous iteration.

\end{itemize}
The former approach is preferable, particularly when the
molecule under consideration is an important shock coolant
(e.g. H$_2$) -- which is not the case of SiO. Furthermore,
the flow time in the SiO emission zone is sufficiently long that the
assumption of local statistical equilibrium is justified.  Accordingly, 
we adopted the latter approach in the present work. It is common to most  applications of the LVG method
to astrophysical problems (e.g. Sch97, Neufeld \& Kaufman 1993) and has the advantage of being less demanding in CPU time. For more significant coolants, such as CO,
$^{13}$CO, and H$_2$O, the rate of cooling was computed in parallel with the shock dynamics, using the cooling functions calculated by \cite{neufeld}, by means of the LVG method.

Note that Equs.~(\ref{eq:matrix}) may be put in a different (matrix) form
by expressing explicitly $\bar J$ as a function of $B_\nu(\mTex)$ and
$I_{\rm c}$. Using the standard relationships between Einstein coefficients,
and the definition of \Tex\ in terms of $n_j$ and $n_{j+1}$, all terms
involving \Tex\ cancel from the equations, leaving radiative
terms which are proportional to the \abeta's (see Goldreich \& Kwan 1974):
\begin{eqnarray}
\label{eq:kwan}
\frac{dn_{i}}{dt} = 
                    & + & n_{i+1} \bar\beta_{i,i+1} \left (A_{i+1,i} +
B_{i+1,i}I_{\rm c} \right )\nonumber \\
		    & - &
n_{i}\left(\bar\beta_{i-1,i}A_{i,i-1}+\bar\beta_{i,i+1}B_{i,i+1}I_{\rm c}+\bar\beta_{i-1,i}B_{i,i-1}I_{\rm c} \right )\nonumber \\
                    & + & n_{i-1} \bar\beta_{i-1,i} B_{i-1,i}I_{\rm c}  \nonumber\\
		    & + & \sum_{\rm coll} \sum_{j \neq i} n_{\rm  coll}
(n_{j}C_{ji} - n_{i}C_{ij}).
\end{eqnarray}

Although the two formulations are equivalent and converge to the same
solution, we have found that inversion of the latter matrix encounters
numerical instabilities and convergence problems at high 
optical depths, possibly due to round--off errors in the (vanishingly small)
\abeta\ terms. In the first formulation, the radiative elements of the
matrix are never zero, even at high opacity, and we obtain much better
convergence and accuracy. Hence we have adopted (\ref{eq:matrix}) in the present
calculations. The ``lambda-iteration'' is terminated when we reach
convergence of the level populations to 1 part in $10^{4}$.

Our code was tested thoroughly against the routine used by Sch97, in the case \abeta = $\beta_\perp$, and against the
web--based online version of RADEX ({\tt www.sron.rug.nl/~vdtak/radex/radex.php}), which uses yet another expression for the escape probability, valid for a
turbulent homogeneous sphere (Van der Tak et al. 2007). Discrepancies never exceeded a few percent.

The MHD shock code provides the physical and chemical profiles (of the temperatures, densities, velocities, and abundances) which are required in order to apply the LVG technique. For SiO, we used the rate coefficients for collisional de-excitation  published by \cite{turner}, for which the collision partner is ground state para--H$_2$. These data are available for rotational quantum numbers $0 \le J \le 20$ and for kinetic temperatures $T = 20$, 40, 70, 100, 150, 200, 250, 300~K and are interpolated to intermediate values of $T$. We made use also of the extrapolated data of \cite{schoier}, which extend to $J = 40$ and $T = 2000$~K. Subsequent calculations, using the rate coefficients of \cite{dayou} for collisions of SiO with para--H$_2$, have shown that the rotational line intensities are insensitive to these collision rates because the lines are formed under conditions which approach LTE. For collisions of SiO with He, we used the rate coefficients of \cite{dayou}, for $J \le 26$ and kinetic temperatures in the range $10 \le T \le 300$~K. At temperatures higher than the maximum for which the rate coefficients were calculated, we assumed that they remain constant. Upwards (excitation) rate coefficients were obtained from detailed balance.

The Einstein A-values, and the rotational constant $B_0 = 21711.967$~MHz (corresponding to $hB_0/k_{\rm B} = 1.042$~K) for the ground
vibrational state of $^{28}$Si$^{16}$O were taken from the NIST
database ({\tt www.nist.gov/data/}).

\subsection{Emergent line intensities and radiation temperatures}
\label{sec:line}

Each layer of the shock wave, of thickness $|\Delta z|$, elementary surface $\Delta S$, and velocity $v_z$, emits the following luminosity (in
erg s\(^{-1}\)) over all directions in a given transition 
$j+1 \rightarrow j$ of frequency $\nu = 2B_0(j+1)$:
\begin{equation}
\label{eq:flux}
F_\nu(v_z) = \bar{\beta} \, {h\nu}\, n_{j+1}A_{j+1,j} \, \mid \Delta z \mid \, \Delta S.
\end{equation}
In order to compute line profiles, however, we need to evaluate the {\it
specific intensity} per unit solid angle, frequency interval, and
projected emitting area (in erg cm\(^{-2}\) s\(^{-1}\) Hz\(^{-1}\)
sr\(^{-1}\)). Following Sch97, we assume that the shock is viewed  face--on,
i.e. in the $z$-direction, normal to the shock front. The intensity is
\begin{equation}
\label{eq:specific}
I_\perp(v_z) = \frac{\beta_\perp }{4\pi} \, h\nu \, n_{j+1}A_{j+1,j} \frac{\mid
\Delta z\mid}{|\Delta \nu_z|},
\end{equation}
where ${|\Delta \nu_z|}$ is the Doppler width of the layer, viewed along the $z$-axis:
\begin{equation}
\label{eq:dnu}
{|\Delta \nu}_z| = \frac{\nu}{c} | \Delta{v_z}| = \frac{\nu}{c} |\Delta
z| \times |{\rm d}v_z/{\rm d}z|. 
\end{equation}
Consequently the intensity becomes
\begin{equation}
\label{eq:intensity}
I_\perp(v_z) = \frac{hc}{4\pi} \frac{\beta_\perp}{\mid {\rm d}v_z/{\rm d}z \mid} n_{j+1}A_{j+1,j}. 
\end{equation}
Noting that \pbeta = $(1-{\rm e}^{-\tau_\perp})$/\ptau\ and using the definition of $\tau$ in
Equ.~(\ref{eq:tau}) as well as the relationships between $A$ and $B$-Einstein
coefficients, one obtains the alternative, simpler expression
\begin{equation}
\label{eq:inu}
I_\perp(v_z) = B_\nu(\mTex) \left (1 - e^{-\tau_\perp} \right )
\end{equation}
which corresponds to the well-known radiative transfer result for a uniform
slab of excitation temperature \Tex\ and opacity \ptau.

In order to compare with actual observed SiO spectra, one needs to
add the cosmic background, and take into account the
{\tt ON-OFF} subtraction applied to radio spectra (to remove atmospheric
noise). The cosmic background at the {\tt ON} position and velocity $v_z$ 
is $B_\nu(T_{\rm bg})$ exp(-\ptau), due to attenuation by the layer, while at the {\tt OFF}
position it is simply $B_\nu(T_{\rm bg})$. The observed intensity is thus
\begin{eqnarray}
\label{eq:iobs}
I_\perp^{\rm obs}(v_z) = {\tt ON - OFF} & = & \left[ B_\nu(\mTex) - B_\nu(T_{\rm bg})\right ] 
\left (1 - e^{-\tau_\perp} \right ) \nonumber \\
                       & = & I_\perp(v_z) \times \left[ 1 - B_\nu(T_{\rm bg})/B_\nu(\mTex)\right].
\end{eqnarray}
Finally, we convert the observed specific intensity of the layer to a
{\it line radiation temperature} $T_{\rm R}$ (in kelvin)
using the definition 
\begin{equation}
\label{eq:trad}
T_{\rm R} \equiv \frac{I^{\rm obs} c^2}{2k_B \nu^2}
\end{equation}
which is standard in radioastronomy. Once radiation temperatures are calculated for each shock layer, and thus
each $v_z$, the line profile is integrated over $v_z$ to yield
integrated line intensities, $T{\rm d}V$, in K km s$^{-1}$.  Note that the
specific intensities, radiation temperatures and $T{\rm d}V$ are valid
for a shock that entirely fills the observing beam. Otherwise, the observed
brightness temperature has to be reduced by an appropriate
``surface filling factor'', $f \approx \Delta S$/(beam area).

\subsection{Line profile at arbitrary inclinations}

In the general case of an arbitrary viewing angle, $\mu = \cos\theta$, 
the term \ptau\ in Equ.~(\ref{eq:inu}) is replaced by the LVG opacity in the chosen
direction, $\tau(\mu) = \tau_\perp/\mu^2$. Therefore
\begin{equation}
\label{eq:incl}
T_{\rm R}(\mu) = {T_{\rm R}}_\perp \frac{\left (1 - {\rm e}^{-\tau_\perp/\mu^2} \right )}
{\left (1 - {\rm e}^{-\tau_\perp} \right )}.
\end{equation}
Thus, $T_{\rm R}$ will be multiplied by
$1/\mu^2$ in the optically thin regime and will be unchanged in the optically thick regime. 
At the same time, the velocity of the layer, $v_z$, is replaced by its
projection along the photon path, $\mu v_z$, so the line profile
becomes narrower. A view which is not face--on will leads to a larger $T{\rm d}V$ for
optically thin lines, and to a smaller $T{\rm d}V$ for optically thick lines
(assuming that the source still fills the beam entirely). A quantitative
evaluation of this effect, for our reference shock model, is presented in Section~\ref{sub:sio_4}.

\section{Fractions of Fe, Si and Mg sputtered from olivine}
\label{appendix_bis}

\begin{figure}
\includegraphics[height=20cm]{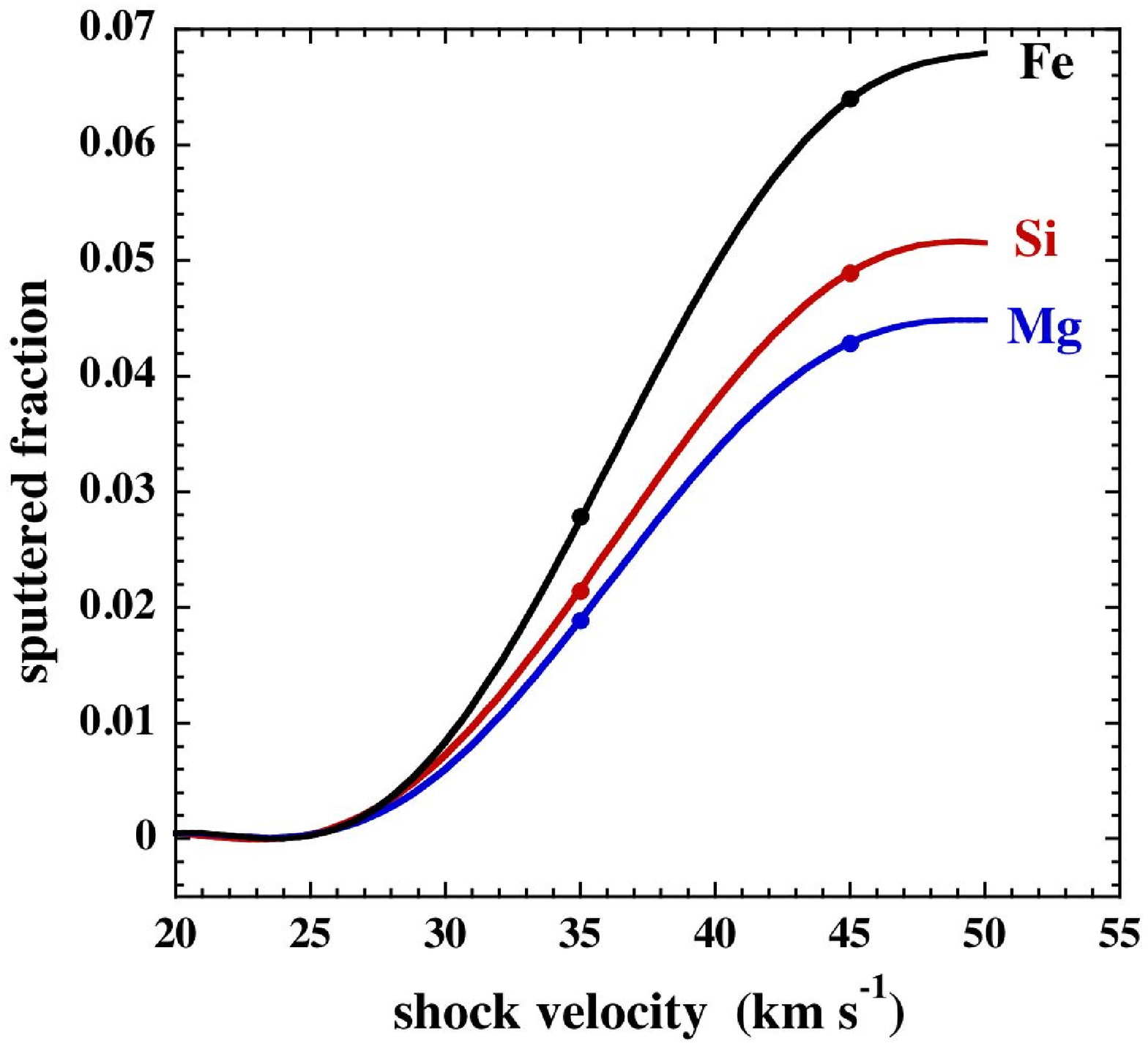}
\caption{Fits of the fractions of Fe, Si and Mg sputtered from olivine grains, as functions of the shock speed. The points on the curves indicate the limiting speeds of steady--state C-type shock waves, for preshock gas densities $n_{\rm H} = 10^{5}$ and $10^{6}$~cm$^{-3}$, with the lower speed corresponding to the higher density. The parameters of the fits to the curves are given in Table~\ref{polyfits}.}
\label{sputt_frac}
\end{figure}

In Section~\ref{sub:grid}, we presented and discussed the sputtering of Fe, Si and Mg from grains composed of olivine (MgFeSiO$_4$). In Fig.~\ref{sputt_frac}, we show the numerical fits to these results, as functions of the shock speed; they are practically independent of the preshock gas density. Also shown in this Figure are the limiting speeds of steady--state C-type shock waves for preshock densities $n_{\rm H} = 10^{5}$ and $10^{6}$~cm$^{-3}$; the limiting speed is lower for the higher density. The limit is associated with the thermal runaway which occurs, owing to the collisional dissociation of H$_2$ within the shock wave. The numerical values of the coefficients of the polynomial fits, of the form $$y({\rm X}) = \sum _{i=0}^5 a_i({\rm X})v_{\rm s}^i,$$ are given in Table~\ref{polyfits}; the shock speed, $v_{\rm s}$, is in km~s$^{-1}$. We have assumed that the magnetic field parameter $b = 1$.

\begin{table}
\caption{The numerical values of the coefficients, $a_i({\rm X})$, of the polynomial fits to the fractions of X $\equiv $ Fe, Si and Mg sputtered from olivine grains. Numbers in parentheses are powers of 10.}
\vspace{1em}
\begin{tabular}{lcccccc}
\hline
  X & $a_0$ & $a_1$ & $a_2$ & $a_3$ & $a_4$ & $a_5$ \\
\hline
\hline
Mg & -0.5322 & 0.09172 & -0.006035 & 0.0001879 & -2.746(-06) & 1.522(-08) \\
Si & -0.4147 & 0.07364 & -0.004953 & 0.0001560 & -2.279(-06) & 1.250(-08) \\
Fe & -0.8579 & 0.14681 & -0.009607 & 0.0002979 & -4.349(-06) & 2.413(-08) \\
\hline
\end{tabular}
\label{polyfits}
\end{table}

\section{Initial gas--phase abundance of O$_2$}
\label{O_2}

\begin{figure}
\includegraphics[height=20cm]{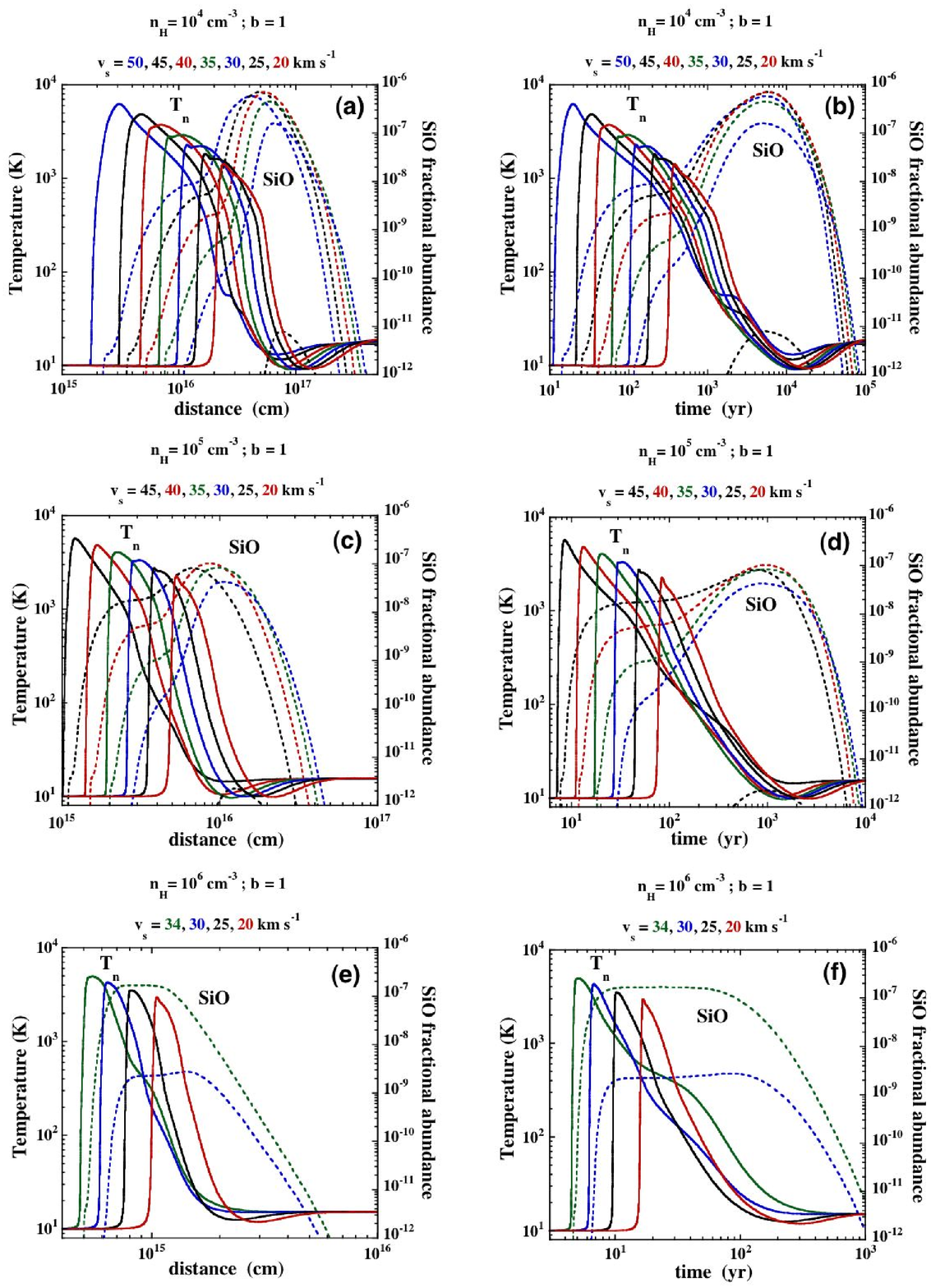}
\caption{As Fig~\ref{sio_1}, but assuming that the initial abundance of O$_2$ ice is negligible (the second of the two scenarios described in Section~\ref{sub:comparison}).}
\label{O2_4}
\end{figure}

\begin{figure}
\includegraphics[height=20cm]{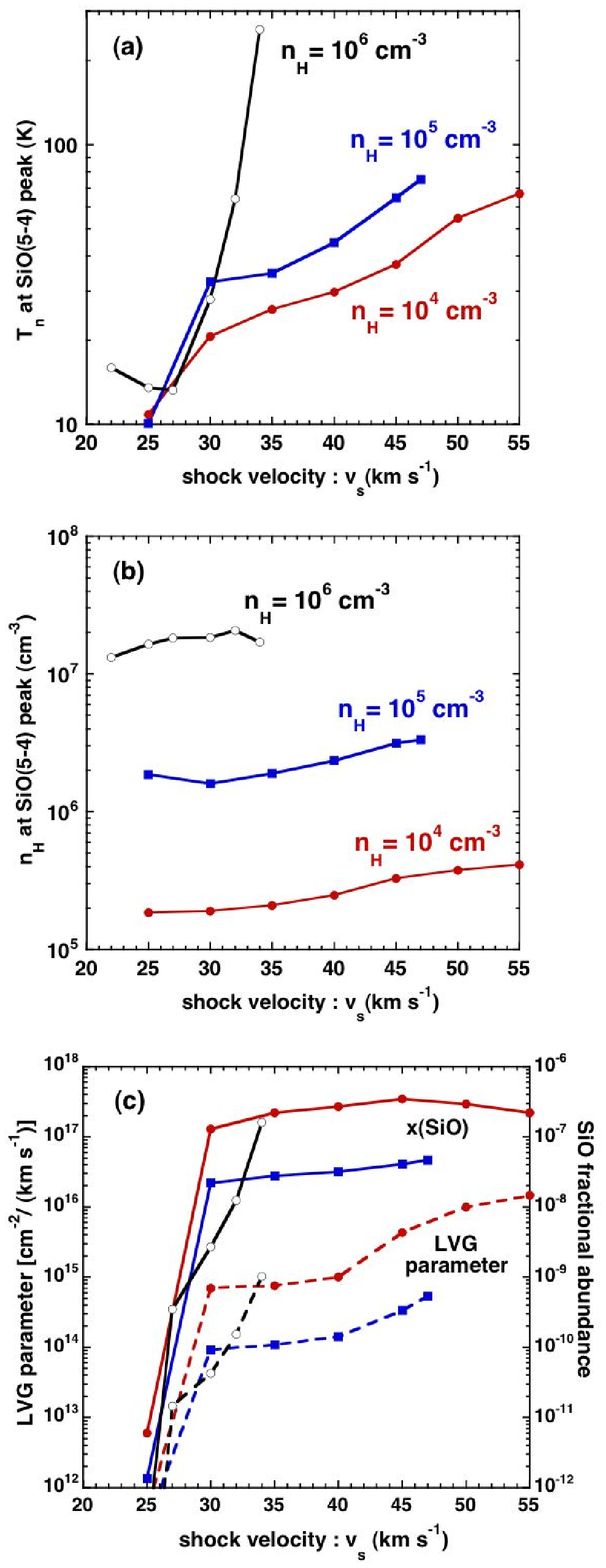}
\caption{As Fig~\ref{phys_cond_2}, but assuming that the initial abundance of O$_2$ ice is negligible (the second of the two scenarios described in Section~\ref{sub:comparison}).}
\label{O2_7}
\end{figure}

\begin{figure}
\includegraphics[height=20cm]{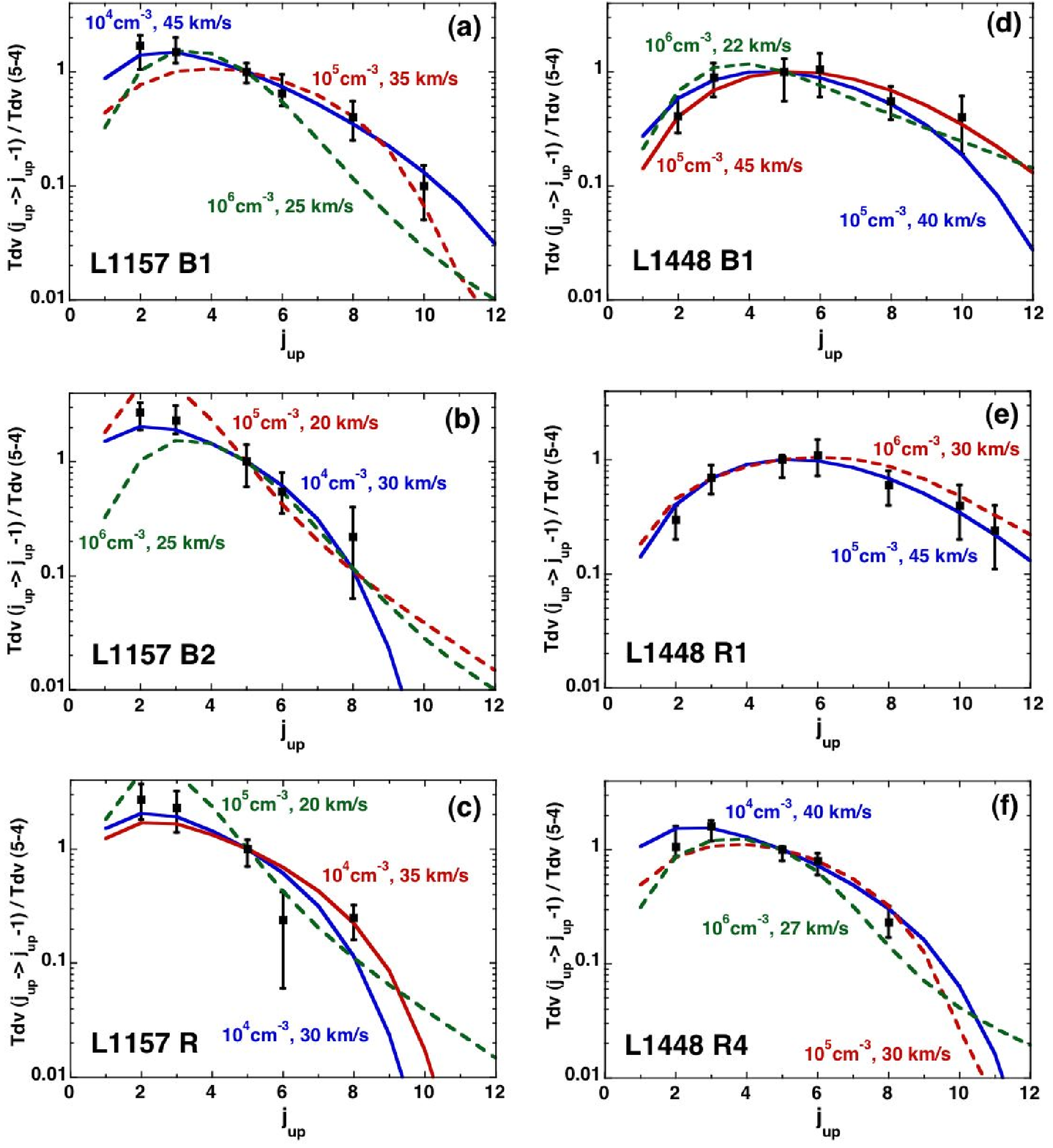}
\caption{As Fig~\ref{profiles}, but assuming that the initial abundance of O$_2$ ice is negligible (the second of the two scenarios described in Section~\ref{sub:comparison}).}
\label{O2_8}
\end{figure}

\begin{figure}
\includegraphics[height=20cm]{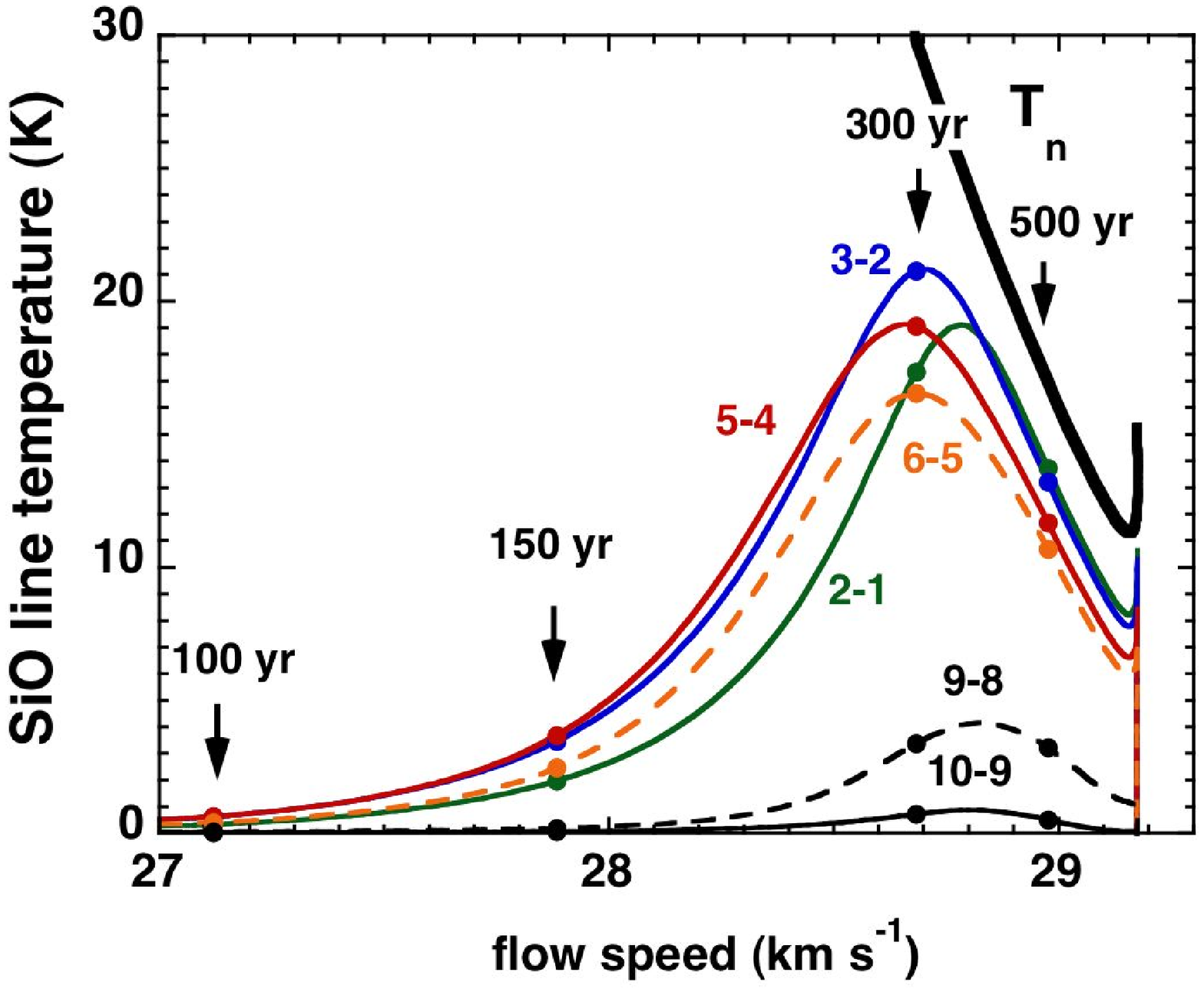}
\caption{As Fig~\ref{sio_rel}, but assuming that the initial abundance of O$_2$ ice is negligible (the second of the two scenarios described in Section~\ref{sub:comparison}).}
\label{O2_12}
\end{figure}

We present, in Figs.~\ref{O2_4}--\ref{O2_12}, the results corresponding to those  in Figs.~\ref{sio_1}, \ref{phys_cond_2}, \ref{profiles} and \ref{sio_rel}, respectively, of the main text, but derived from our secondary grid of models, in which $n({\rm O}_2)/n_{\rm H} = 1.0\times 10^{-7}$ initially in the gas phase but the excess
oxygen is in form of H$_2$O ice, rather than O$_2$ ice as in our
primary grid. We recall that the initial fractional abundance of O$_2$ in chemical equilibrium is $n({\rm O}_2)/n_{\rm H} \approx 10^{-5}$.

In these models, O$_2$ is never abundant in the gas phase, and Si
oxidation occurs only in reaction~\ref{eqs2}, with OH. Thus, SiO formation is
less efficient at intermediate shock speeds and not
significant at $v_{\rm s} = 25$~km~s$^{-1}$.

\end{document}